\documentclass[a4paper,fleqn,usenatbib]{mnras}

\usepackage{newtxtext,newtxmath}

\usepackage[T1]{fontenc}
\usepackage{ae,aecompl}

\usepackage{graphicx}	
\usepackage{amsmath}	
\usepackage{amssymb}	

\title[CMEs on cool stars I]{Stellar Coronal Mass Ejections I. Estimating occurrence frequencies and mass-loss rates}

\author[P. Odert et al.]{
P. Odert,$^{1,2}$\thanks{E-mail: petra.odert@oeaw.ac.at}
M. Leitzinger,$^{1}$
A. Hanslmeier$^{1}$
and H. Lammer$^{2}$
\\
$^{1}$Institute of Physics/IGAM, University of Graz,
Universit\"atsplatz 5, A-8010 Graz, Austria\\
$^{2}$Space Research Institute, Austrian Academy of Sciences,
Schmiedlstra{\ss}e 6, A-8042 Graz, Austria
}

\date{Accepted XXX. Received YYY; in original form ZZZ}

\pubyear{2017}

\begin{document}
\label{firstpage}
\pagerange{\pageref{firstpage}--\pageref{lastpage}}
\maketitle

\begin{abstract}
Stellar coronal mass ejections (CMEs) may play an important role in mass- and angular momentum loss of young Sun-like stars. If occurring frequently, they may also have a strong effect on planetary evolution by increasing atmospheric erosion. So far it has not been possible to infer the occurrence frequency of stellar CMEs from observations. Based on their close relation with flares on the Sun, we develop an empirical model combining solar flare--CME relationships with stellar flare rates to estimate the CME activity of young Sun-like and late-type main-sequence stars. By comparison of the obtained CME mass-loss rates with observations of total mass-loss rates, we find that our modeled rates may exceed those from observations by orders of magnitude for the most active stars. This reveals a possible limit to the extrapolation of such models to the youngest stars. We find that the most uncertain component in the model is the flare--CME association rate adopted from the Sun, which does not properly account for the likely stronger coronal confinement in active stars. Simple estimates of this effect reveal a possible suppression of CME rates by several orders of magnitude for young stars, indicating that this issue should be addressed in more detail in the future.
\end{abstract}

\begin{keywords}
stars: activity -- stars: mass-loss -- stars: late-type
\end{keywords}



\section{Introduction}
\label{sec:intro}
Flares and Coronal Mass Ejections (CMEs) belong to the most energetic activity phenomena on the Sun. Whereas it is well established that other Sun-like and cooler stars generate flares \citep[e.g.][]{Gershberg75, Maehara12, Balona15, Davenport16}, the situation is different for stellar CMEs, since they are more difficult to observe. Obtaining a better understanding of CMEs on main-sequence FGKM stars is of special importance, because these stars could potentially host habitable planets. Since planetary habitability depends, among other factors, also on the conditions and stability of an atmosphere, all processes contributing to atmospheric erosion may severely restrict the habitability of a given planet, even if it orbits within the habitable zone of its host star. Young FGKM stars rotate faster than the Sun and have correspondingly higher levels of magnetic activity, leading to enhanced X-ray--UV (XUV) emission and frequent, powerful flares. Recent discoveries include superflares with bolometric energies up to $10^{37}$\,erg \citep{Maehara12, Wu15}, some even occurring on slowly rotating Sun-like stars \citep{Nogami14}. Enhanced radiation levels, either constant or temporarily variable during flares, affect planetary atmospheres by heating, ionization, and dissociation, leading to increased thermal escape \citep{Watson81, Erkaev13, Lammer14}. In addition, strong stellar winds and frequent CME impacts may enhance non-thermal loss processes. Ion pick-up loss rates could be high in close orbits around young stars, which is especially relevant for planets in the habitable zones (HZs) of M~dwarfs, since they have likely weak magnetic moments due to tidal locking \citep{Lammer07a, Khodachenko07a}. Strong planetary magnetic fields could reduce the effect of this process and also moderate atmospheric losses due to thermal escape \citep{Khodachenko07, Khodachenko15a}, but efficient polar winds could be generated if the atmospheres are highly ionized by the stellar XUV flux \citep{Cohen12, Garcia-Sage17, Airapetian17}. Simulations of the planetary system orbiting the M~dwarf TRAPPIST-1 indicate that stellar and planetary magnetic field lines are connected and the stellar wind may flow directly onto the atmospheres of the planets, which could lead to strong atmospheric erosion \citep{Garraffo17}.

On the Sun, CMEs are often associated with flares and are thought to be caused by the same underlying physical processes \citep{Priest02, Compagnino17}. However, not all flares are accompanied by CMEs and not all CMEs by flares. Flares without CMEs are typically shorter in duration and less energetic, whereas longer and more energetic flares are more likely to be associated with CMEs \citep{Yashiro06, Compagnino17}. Although there are also CMEs which are apparently not related to flares or other low coronal signatures (often termed ``stealth CMEs''), these are typically less massive, less wide, and on average slower than normal CMEs \citep{DHuys14}. However, \citet{Alzate17} recently found that all stealth CMEs studied by \citet{DHuys14} turned out to be artifacts of observational and data processing limitations. With new image processing techniques, low coronal signatures could be identified for all events, indicating that this type of CMEs likely does not comprise a physically distinct population.

CMEs are also often related to erupting prominences (EPs) which are believed to form the CME cores in the frequently encountered three-part structure consisting of a bright core, a dark cavity and a narrow, bright leading edge \citep{Forbes00}. About 70\% of EPs are associated with CMEs and vice versa \citep{Munro79, Gopalswamy03}. EPs not associated with CMEs have mostly non-radial motion or are stalled by the overlying magnetic field \citep{Gopalswamy15e}. It is important to note that the association rates between these phenomena can vary significantly between studies because of different samples, projection and selection effects, as well as different detection limits of the instruments needed to identify these different phenomena \citep[e.g.][]{Burkepile04, Cremades15}. The occurrence rate of solar CMEs varies in the course of the solar cycle from about one every few days in minimum to a few per day in maximum. The associated mass-loss rate ranges from about 1.5 to $4.8\times10^{-16}\,M_{\sun}\,\mathrm{yr^{-1}}$ from minimum to maximum \citep{Vourlidas10}, i.e. a few per cent of the average mass-loss rate from the solar wind \citep[${\sim}2\times10^{-14}\,M_{\sun}\,\mathrm{yr^{-1}}$;][]{Wang98}. CME masses can be up to several $10^{16}$\,g, whereas their velocities range from a few 100 to a few 1000\,km\,s$^{-1}$ \citep{Webb12}.

It has been hypothesized that active stars with their frequent and powerful flares could also have correspondingly numerous energetic CMEs. These could be an important contribution to stellar mass- and angular momentum loss; therefore, they could strongly influence the evolution of young stars \citep{Aarnio12, Drake13, Osten15a, Cranmer17}. Observations of CMEs on other stars are challenging because the close stellar environment cannot be spatially resolved like in solar observations. Up to now, there are only a few cases in which stellar mass ejections have been observed. Several events were detected as transient blue-shifted emission components in spectral lines during flares \citep{Houdebine90, Gunn94a, Guenther97, Fuhrmeister04, Leitzinger11a, Vida16}. Other indications of ejected mass during stellar flares stem from transient absorptions seen in UV or X-rays, indicating excess neutral material rising above the flaring region and obscuring the emission \citep{Giampapa82a, Haisch83, Ambruster86, Doyle88a, Wheatley98, Favata99, Pandey12}. However, such observations are more difficult to interpret because of the lack of information on the plasma velocity. Mass estimates of observed stellar CME events range from $10^{15}$ to $10^{19}$\,g and line-of-sight velocities from ${\sim}100$ to 5800\,km\,s$^{-1}$. The masses are comparable to those estimated for stellar prominences \citep{CollierCameron89, CollierCameron90, Dunstone06a, Leitzinger16}, which can be observed as transient features in the broadened Balmer line profiles of fast rotating young stars. Prominences consist of cool chromospheric material and often form the cores of CMEs if they erupt, which explains why stellar mass ejection events can be observed in the Balmer lines. Another observational attempt to detect stellar CMEs is the search for radio type~II bursts, which are related to propagating shock fronts and are therefore often associated with CMEs on the Sun \citep{Reiner01a, Gopalswamy05}. However, up to now, no radio bursts of this type were detected on other stars \citep{Leitzinger08, Leitzinger10, Boiko12, Crosley16}.

Previous studies aimed to establish correlations between solar flare energies and CME parameters to estimate the possible occurrence frequencies, as well as associated mass-, energy- and angular momentum loss rates on young stars \citep{Aarnio12, Drake13, Osten15a}. \citet{Leitzinger14} used a similar approach to interpret the non-detection of mass ejections in young open cluster stars. Recently, \citet{Cranmer17} attempted to infer stellar CME mass-loss rates from surface-averaged magnetic fluxes. However, the question remains how reliable such estimates based on the extrapolation of the solar relations to stars with much higher activity levels are. In this study, we describe an empirical model similar to those mentioned before to address this question in more detail. In section~\ref{sec:model}, we describe the model, including similarities and differences to previous approaches. In section~\ref{sec:comp}, we compare the estimated CME occurrence and mass-loss rates of active Sun-like stars with existing observations of stellar mass-loss rates. Possible limitations of such extrapolations will be discussed in section~\ref{sec:disc} and we summarize our findings in section~\ref{sec:sum}.

\section{Model description}\label{sec:model}
Here, we develop an empirical model with the aim to estimate stellar CME rates from combining stellar flare rates and observed flare--CME relationships from the Sun. This approach is similar to previous studies \citep{Aarnio12, Drake13, Osten15a}, but differs in that we are using purely observational relations, similar as in \citet{Leitzinger14} and \citet{Odert16}.

The differential distribution of flare-related CMEs as a function of their masses $M$ can generally be written as
\begin{equation}\label{eq:dndm}
    \frac{dN}{dM} = \frac{dN}{dE_*} \frac{dE_*}{dE_{\sun}}\frac{dE_{\sun}}{dM}P(M),
\end{equation}
where $N$ is the number of events per unit time and $E_{*/\sun}$ are the stellar/solar flare energies. The first term on the right hand side is the differential distribution of stellar flares as a function of their energy. The second term accounts for the different energy bands of solar and stellar observations. The third term describes the relation between flare energy and CME mass on the Sun. The fourth term describes the probability that a flare is associated with a CME (i.e., the CME-flare association rate). We estimate each of these terms below.

\subsection{Stellar flare rates}
The differential distribution of both solar and stellar flares as a function of their energy is well represented by a power law
\begin{equation}\label{eq:dnde}
    \frac{dN}{dE} = k_E E^{-\alpha}
\end{equation}
with index $\alpha$ and constant $k_E$. Typical values of $\alpha$ range from 1.4 to 2.7 in both solar and stellar flare observations \citep[e.g.][and references therein]{Guedel03}. The values show some dependence on the wavelength range and the covered flare energy range. Despite the similarities of $\alpha$ found in solar and stellar observations, the constant $k_E$ generally differs, because stars with a higher activity level have more flares of a given energy than less active stars \citep{Gershberg89}. \citet{Audard00} found that the daily number of stellar X-ray flares with energies $E_{X}{>}10^{32}$\,erg is closely correlated with the total stellar X-ray luminosity $L_X$ via
\begin{equation}\label{eq:aud}
    \log N(E>10^{32}) [\mathrm{d}^{-1}] = a + b \log L_{X},
\end{equation}
with $a=-26.7\pm2.9$, $b=0.95\pm0.1$, and $L_X$ given in erg\,s$^{-1}$. This scaling holds for stars with $L_X{\sim}10^{27}$ to $10^{31}$\,erg\,s$^{-1}$. From Eq.~\ref{eq:dnde}, one can deduce an expression for the cumulative flare distribution
\begin{equation}
    N(>E) = \int_{E}^\infty \frac{dN}{dE'}\ dE' = \frac{k_{E}}{\alpha-1}E^{1-\alpha},
\end{equation}
representing the number of flares above energy $E$. Comparing with Eq.~\ref{eq:aud}, one can then express $k_E$ as a function of $\alpha$ and $L_X$ as
\begin{equation}\label{eq:kx}
    k_E = (\alpha-1) 10^{32(\alpha-1)+a} L_X^b.
\end{equation}
This scaling allows to estimate the X-ray flare distribution of any Sun-like main-sequence star simply by its measured $L_X$ as a function of the flare power law index $\alpha$.

\subsection{Conversion of energy bands}
It is important to note that stellar X-ray observations are typically performed in different bands compared to solar observations. Therefore, some conversion to relate the flare energies in solar and stellar bands has to be applied. \citet{Audard00} used a constant count-to-flux conversion factor for both flares and quiescent emission corresponding to the combined X-ray and EUV (XUV; 0.01--10\,keV; 0.1--124\,nm) range. However, they note that most of the energy release of their target stars occurs within 0.1--5\,keV (0.25--12.4\,nm), comparable to the 0.1--2\,keV band of \textit{ROSAT}. Comparison with \textit{ROSAT} observations \citep{Schmitt04} of the stars in their sample shows that their given XUV luminosities are comparable to the X-ray values. In some cases, they are up to 0.1--0.2\,dex higher, which is, however, comparable to the scatter between different \textit{ROSAT} observations of the same star at different times. Thus, both the XUV luminosities and flare energies given in \citet{Audard00} should be comparable to broad band stellar X-ray data.

On the other hand, all solar relations used here relate to measurements from the \textit{Geostationary Operational Environmental
Satellite (GOES)} performed in the 0.1--0.8\,nm (1.55--12.4\,keV) band, commonly used to classify solar flares. To convert between solar and stellar flare energies, we adopt the relation between flare fluences\footnote{Fluence (J\,m$^{-2}$) is converted to energy at the Sun (erg) with the factor $1.406\times10^{30}$, which corresponds to \mbox{$10^{7}\times2\pi(\mathrm{AU})^{2}$\,erg}, assuming uniform radiation into the visible hemisphere of the Sun \citep{Woods06}.} $F_{GOES}$ in the \textit{GOES} band and $F_{XPS}$ in the XUV Photometer System (\textit{XPS}) (0.1--27\,nm) band found by \citet{Woods06}, $F_{XPS}=63 \times F_{GOES}^{0.8}$. Although the \textit{XPS} band is not exactly equal to that of the stellar flare study from \citet{Audard00}, it should be a reasonable proxy. Converting fluence to energy and setting $E_{XPS}{\approx}E_X$, we find
\begin{equation}\label{eq:estar}
E_X=\eta E_{GOES}^{\xi}
\end{equation}
with $\eta=6.7\times10^7$ and $\xi=0.8$. Note that $E_{XPS}$ is within less than a factor of two compared to the total 0--190\,nm emission in several large solar flares \citep{Woods06}, indicating that it is likely a good proxy for the flare energies in the XUV range.

\subsection{Solar flare energy and CME mass}
On the Sun, flare energies and the masses of their associated CMEs are correlated \citep{Aarnio12, Drake13, Takahashi16, Compagnino17}. This is also physically plausible because the energies of CMEs and the bolometrically radiated flare energies are of similar magnitude \citep{Emslie12}. The CME masses are related with the \textit{GOES} X-ray energies of their associated flares via a power law
\begin{equation}\label{eq:me}
    M = \mu E_{GOES}^\beta
\end{equation}
with parameters $\mu=(2.7\pm1.2)\times10^{-3}$, $\beta=0.63\pm0.04$ \citep{Aarnio12} and $\mu=10^{-1.5\mp0.5}$, $\beta=0.59\pm0.02$ \citep{Drake13}, respectively. The masses for both parameter sets are rather similar and deviate only for the highest masses by a factor of about two. This is negligible compared to the intrinsic spread of about an order of magnitude about this relation. Hereafter we adopt the parameters of \citet{Drake13}.

\subsection{CME--flare association rate}
The solar CME--flare association rate increases with flare energy.  \citet{Yashiro06} obtained an expression as a function of solar \textit{GOES} flare fluence which can be written as $P(F_{GOES})=0.371(\log F_{GOES}+3.3)$. This relation is in good agreement with the more recent studies of \citet{Yashiro09}. By converting fluence to energy as before and using Eq.~\ref{eq:me}, we can express $P$ as a function of CME mass
\begin{equation}\label{eq:pm}
    P(M) = c+d\log M
\end{equation}
with $c=-9.02$ and $d=0.63$. Equation~\ref{eq:pm} is valid between $M_1=8.5\times10^{15}$\,g where $P(M_1)=1$ and $M_0=2.2\times10^{14}$\,g where $P(M_0)=0$. For $M>M_1$, $P(M_1)=1$, i.e. all flares are associated with CMEs. Note that Eq.~\ref{eq:pm} is strictly not valid for $P(M) \rightarrow 0$ \citep{Yashiro06}. However, this lower cut-off is adopted here because the CME--flare association is not well established for weak flares due to the sensitivity limits of solar instruments and selection effects. Thus, the inferred CME occurrence rates estimated here represent a lower limit. On the other hand, weak flares are typically associated with less massive CMEs (Eq.~\ref{eq:me}). Therefore, the computed mass-loss rates (section~\ref{sec:mdotcme}) will not be significantly by this approach. The mass limits $M_0$ and $M_1$ correspond to X-ray flare energies of $2\times10^{29}$ and $2.9\times10^{31}$\,erg, as well as to \textit{GOES} flare energies of $7.1\times10^{26}$ and $3.5\times10^{29}$\,erg, respectively.

\subsection{Inferred stellar CME distribution}
We now determine $dN/dM$ (Eq.~\ref{eq:dndm}) using Eqs.~\ref{eq:dnde}, \ref{eq:estar}, \ref{eq:me}, and \ref{eq:pm} by evaluating all required terms and expressing them as a function of CME mass $M$. This yields the stellar CME distribution
\begin{equation}\label{eq:dndm2}
\frac{dN}{dM} = P(M)k_M M^{-\gamma},
\end{equation}
where we have defined
\begin{equation}\label{eq:km}
k_M=k_E\frac{\xi}{\beta}\eta^{1-\alpha}\mu^{\gamma-1},\ \gamma=1-\frac{\xi(1-\alpha)}{\beta},
\end{equation}
with $k_E$ from Eq.~\ref{eq:kx} and $P(M)$ from Eq.~\ref{eq:pm}. For CME masses larger than $M_1$, Eq.~\ref{eq:dndm2} yields a power law similar to the flare distribution (Eq.~\ref{eq:dnde}). With Eq.~\ref{eq:dndm2}, one can deduce the CME occurrence rate for any star given its X-ray luminosity $L_X$ and flare index $\alpha$. The magnitude of the distribution mainly depends on $L_X$, because $k_M$ is a function of $L_X$, whereas the slope is determined by $\alpha$ if all other scaling law parameters are fixed. A typical range of $\alpha=1.5$ to 2.5 translates to a range of $\gamma=1.67$ to 3.03.

\begin{figure}
\centering
\includegraphics[width=\columnwidth]{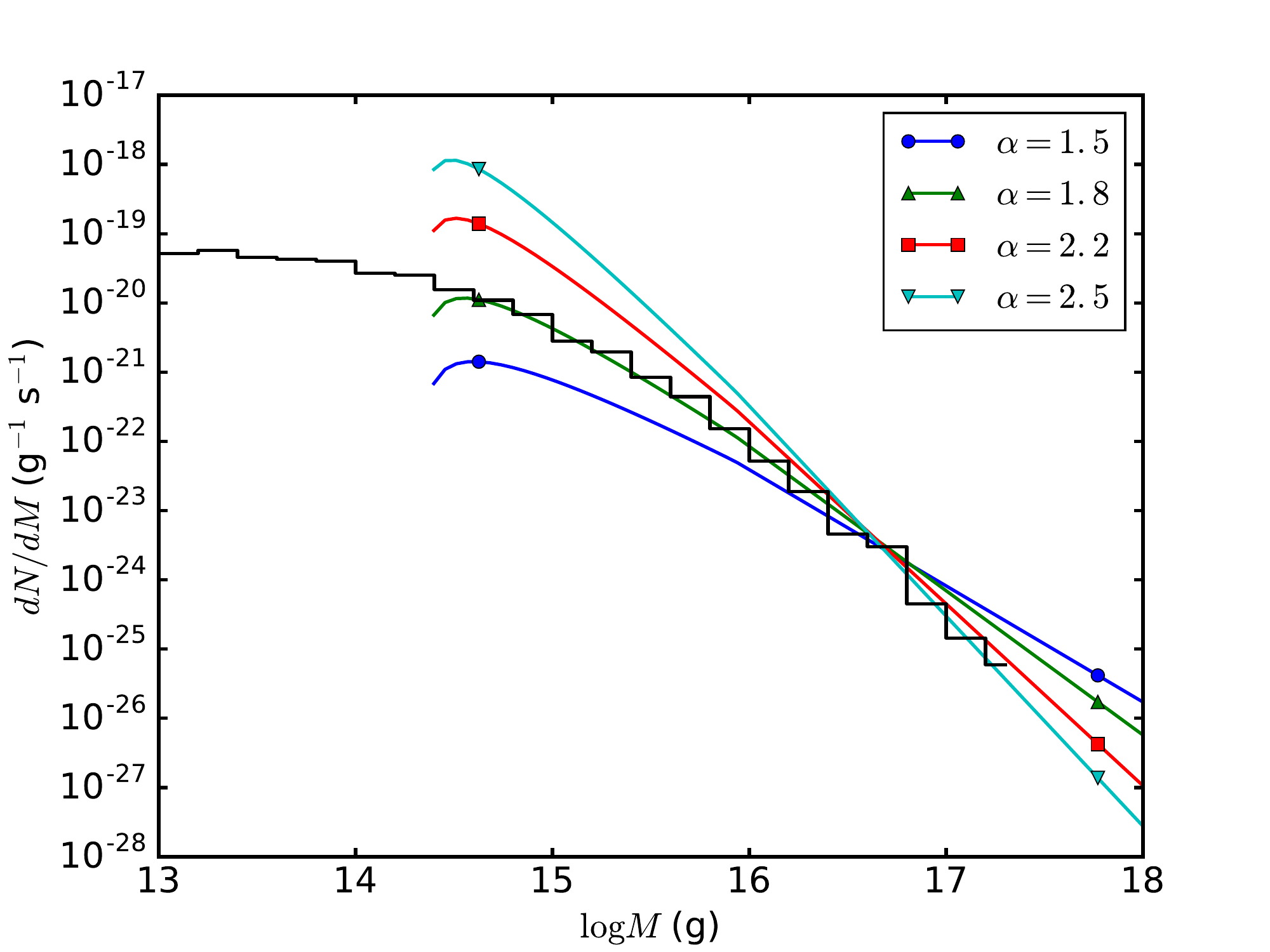}
\caption{Differential CME distribution for a star with $\log L_X=26.5\,\mathrm{erg\,s^{-1}}$ (comparable to solar minimum conditions) and four values of $\alpha$ covering the typically observed range. The observed solar CME distribution \citep{Aarnio12} is shown in black.}
\label{fig:dndm}
\end{figure}

In Fig.~\ref{fig:dndm}, we show the inferred differential CME distribution of a star with $\log L_X=26.5\,\mathrm{erg\,s^{-1}}$ and four values of $\alpha$. It is obvious that smaller values of $\alpha$ yields a higher number of massive CMEs and a lower number of low-mass CMEs compared to larger values of $\alpha$. Using higher values of $L_X$ yields similar distributions, but simply shifted upwards along the $y$-axis. We compare the results with the observed solar differential CME distribution, which was found to follow a power law for the high mass tail with an index $\gamma=2.1$ \citep{Aarnio12}. Using Eq.~\ref{eq:dndm2}, we can reproduce this slope using a broad-band X-ray flare index $\alpha=1.8$. This index is in good agreement with solar flare observations \citep{Hudson91} and the distribution of thermal energies \citep{Aschwanden15}. Thus, we conclude that the power law indices used in the individual scalings entering Eq.~\ref{eq:dndm2} are reasonable. For the highest masses, the observed slope is slightly steeper than predicted and follows the $\alpha=2.5$ curve. This may be due to either the exponential cut-off in flare energies discussed in section~\ref{sec:disc}, i.e. a steepening of the real flare distribution close to the largest flare energies, or the small number of observed CME events in these bins. The adopted X-ray luminosity of $\log L_X=26.5\,\mathrm{erg\,s^{-1}}$ is close to the value estimated for solar minimum conditions in the \textit{ROSAT} band \citep{Peres00}.

\subsection{Occurrence rates}
\begin{figure}
\centering
\includegraphics[width=\columnwidth]{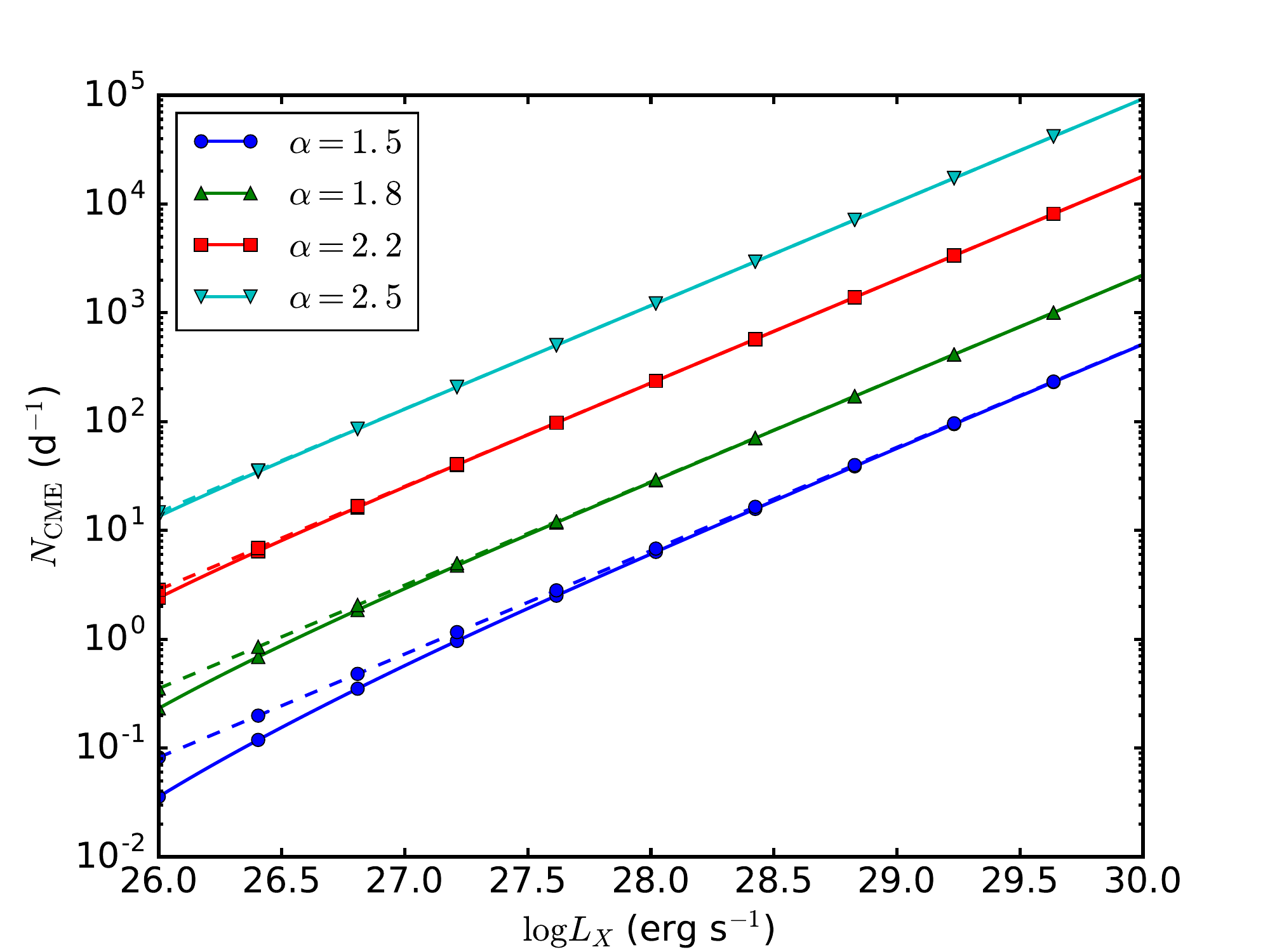}
\caption{Total number of CMEs per day as a function of $L_X$ and $\alpha$. Solid lines indicate rates with integration up to a maximum CME mass, whereas dashed lines show the occurrence rates derived from the cumulative distribution.}
\label{fig:ncme}
\end{figure}

To obtain the occurrence rate of CMEs above a certain mass $M$, one can construct the cumulative distribution from Eq.~\ref{eq:dndm2},
\begin{equation}\label{eq:ncum}
\begin{aligned}
N(>M) &= \int_{M}^{\infty}\frac{dN}{dM'}\ dM'= & \\
&= \begin{cases}
\frac{k_M}{\gamma-1}M^{1-\gamma} & \text{for}\ M \geq M_1 \\
\frac{k_M d}{(\gamma-1)^2\ln(10)} \left(M^{1-\gamma}-M_1^{1-\gamma}\right)+ \\
\quad+\frac{k_M}{\gamma-1}M^{1-\gamma}P(M) & \text{for}\ M_0<M<M_1. \\
\end{cases}
\end{aligned}
\end{equation}
The cumulative number evaluated at the minimum considered CME mass $M_0$ represents approximately the total number of CMEs per unit time. However, the cumulative distribution slightly overestimates the total CME rate because the integration goes to infinity. This overestimate is higher for smaller $\alpha$. However, there likely is an upper limit to possible CME masses for any given star, similar to the maximum possible flare energy \citep{Aulanier13}. Thus, in order to obtain a better estimate of the total CME rate, we integrate Eq.~\ref{eq:dndm2} between $M_0$ and a maximum expected CME mass $M_\mathrm{max}$,
\begin{equation}\label{eq:ncme}
\begin{aligned}
N &= \int_{M_0}^{M_\mathrm{max}}\frac{dN}{dM}\ dM = & \\
&= \begin{cases}
\frac{k_M}{\gamma-1}\left[\frac{d}{\ln(10)(\gamma-1)}\left(M_0^{1-\gamma}-M_1^{1-\gamma}\right)-M_\mathrm{max}^{1-\gamma}\right]\\
\quad\text{for}\ M_\mathrm{max} \geq M_1 \\
\frac{k_M}{\gamma-1}\left[\frac{d}{\ln(10)(\gamma-1)}\left(M_0^{1-\gamma}-M_\mathrm{max}^{1-\gamma}\right)-M_\mathrm{max}^{1-\gamma}P(M_\mathrm{max})\right]\\
\quad\text{for}\ M_0<M_\mathrm{max}<M_1 \\
0\quad\text{for}\ M_\mathrm{max}\leq M_0. \\
\end{cases}
\end{aligned}
\end{equation}
This result is valid for $\gamma\neq1$. Since $\gamma=1$ would require $\alpha=1$ (cf. Eq.~\ref{eq:km}), which is lower than observed for both the Sun and stars, we only give the solution above. We estimate $M_\mathrm{max}$ from the maximum expected flare energies using Eqs.~\ref{eq:me} and \ref{eq:estar}, where the maximum flare energies are estimated as $E_\mathrm{max} \approx 10^{4.5} L_X$. This corresponds roughly to the largest observed flare energies in the stellar sample of \citet{Audard00}, as a function of their X-ray luminosities. This is also comparable to the value of $10^4L_X$ adopted in the study of \citet{Drake13}. Converting this scaling to CME masses yields
\begin{equation}
\begin{aligned}
\log M_\mathrm{max} &= \log\left[\mu\left(\frac{10^{4.5}}{\eta}\right)^{\beta/\xi}\right] + \frac{\beta}{\xi}\log L_X \approx \\
&\approx -3.955 + 0.7375\log L_X.
\end{aligned}
\end{equation}
This scaling yields maximum possible CME masses in the order of $10^{15}$ to $10^{18}$\,g for stars with $L_X{\sim}10^{26}$ to $10^{30}\,\mathrm{erg\,s^{-1}}$. Using Eq.~\ref{eq:ncme}, we calculate the total number of CMEs per day as a function of $L_X$ and $\alpha$ (Fig.~\ref{fig:ncme}), in comparison to the cumulative rates from Eq.~\ref{eq:ncum}. The occurrence rate increases with both $L_X$ and $\alpha$. One can see that there is a very large spread for our adopted range of $\alpha$ by more than two orders of magnitude. This indicates that such a scaling is only useful for stars for which $\alpha$ has been determined with high accuracy. The cumulative rate generally gives a good approximation to that calculated with $M_\mathrm{max}$ for most of the parameter space. It predicts slightly higher rates for stars with low $L_X$ and small $\alpha$.

\subsection{CME mass- and energy loss rates}\label{sec:mdotcme}
\label{lossrates}
\begin{figure}
\centering
\includegraphics[width=\columnwidth]{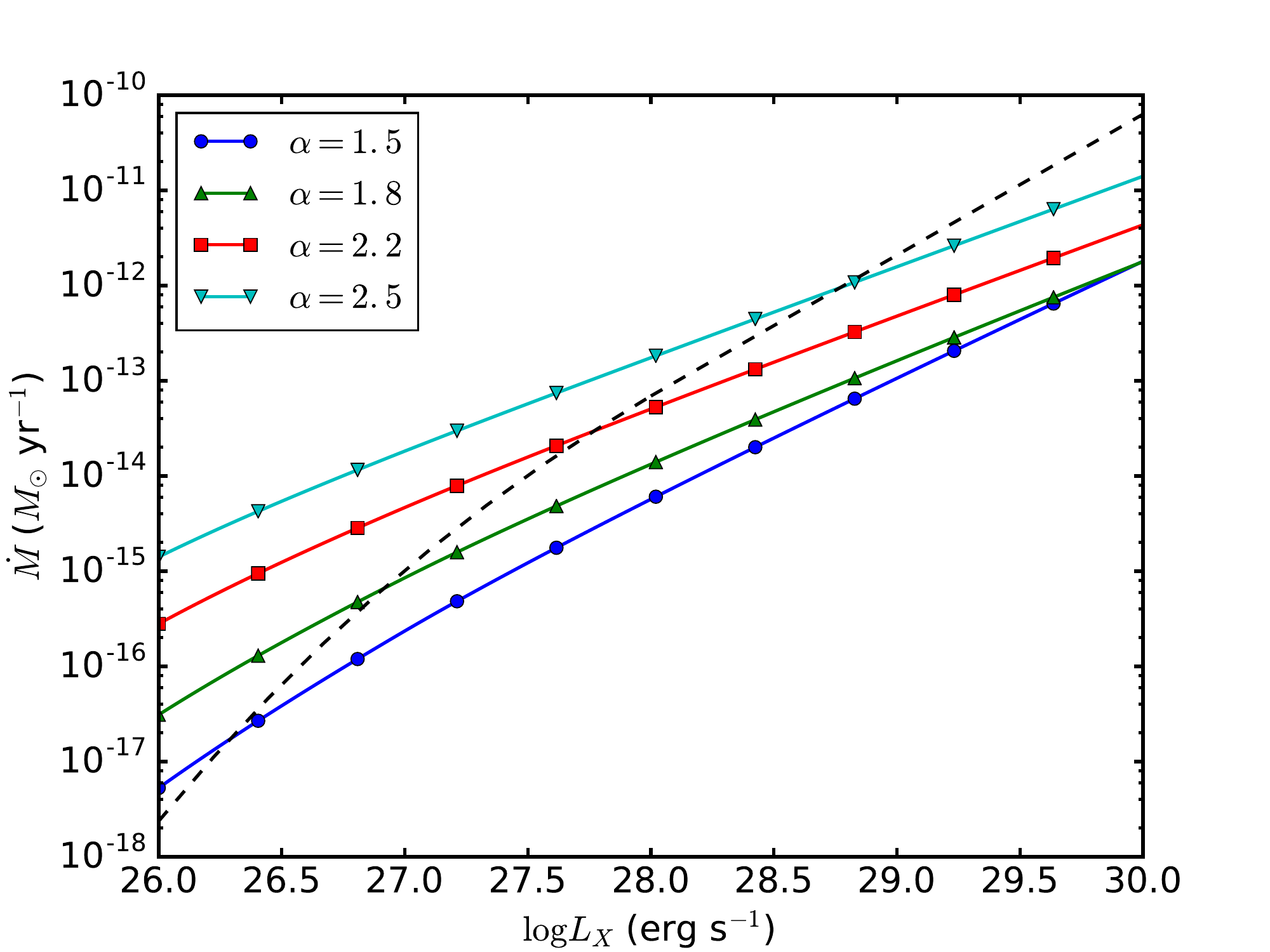}
\caption{CME mass-loss rates as a function of $L_X$ and $\alpha$. The relation from \citet{Drake13} is shown as a dashed line; it has almost negligible dependence on $\alpha$.}
\label{fig:mdot}
\end{figure}

The mass-loss rate due to CMEs can be calculated via
\begin{equation}\label{eq:mdot1}
\begin{aligned}
\dot{M} &= \int^{M_\mathrm{max}}_{M_0} M \frac{dN}{dM} dM = \int^{M_\mathrm{max}}_{M_0} P(M) k_M M^{-\gamma+1} dM \\
&= \begin{cases}
\frac{k_{M}}{\gamma-2}\left(\frac{d}{(\gamma-2)\ln(10)}\left(M_0^{2-\gamma}-M_1^{2-\gamma}\right)-M_\mathrm{max}^{2-\gamma}\right)\\
\quad\text{for}\ M_\mathrm{max} \geq M_1 \\
\frac{k_{M}}{\gamma-2}\left(\frac{d}{(\gamma-2)\ln(10)}\left(M_0^{2-\gamma}-M_\mathrm{max}^{2-\gamma}\right)-M_\mathrm{max}^{2-\gamma}P(M_\mathrm{max})\right)\\
\quad\text{for}\ M_0<M_\mathrm{max}<M_1 \\
0\quad\text{for}\ M_\mathrm{max}\leq M_0 \\
\end{cases}
\end{aligned}
\end{equation}
if $\gamma\neq2$. For $\gamma=2$ (corresponding to $\alpha{\sim}1.74$), the solutions are
\begin{equation}\label{eq:mdot2}
\begin{aligned}
\dot{M} &= \begin{cases}
\frac{k_M}{2}\left[2\ln(M_\mathrm{max})+(c-1)\ln(M_1)-c\ln(M_0)\right]\\
\quad\text{for}\ M_\mathrm{max} \geq M_1\\
\frac{k_M}{2}\left[\ln(M_\mathrm{max})\left(c+P(M_\mathrm{max})\right)-c\ln(M_0)\right]\\
\quad\text{for}\ M_0<M_\mathrm{max}<M_1\\
0\quad\text{for}\ M_\mathrm{max} \leq M_0.\\
\end{cases}
\end{aligned}
\end{equation}
The resulting CME mass-loss rates are shown in Fig.~\ref{fig:mdot}. The mass-loss rates increase with $L_X$ and $\alpha$. As a comparison, we also plot the relation derived by \citet{Drake13}. Their result shows a steeper dependence on $L_X$ than ours, and their dependence on $\alpha$ is almost negligible. Despite being quantitatively similar to our results for intermediate $L_X$, their predicted mass-loss rates are lower (higher) for less (more) active stars. These differences are likely caused by the different method used to calculate $\dot{M}$. Specifically, we find that their weak dependence on $\alpha$ results from their approach to calculate $k_E$, by setting the flare power in the integration range equal to $L_X$.

\begin{figure}
\centering
\includegraphics[width=\columnwidth]{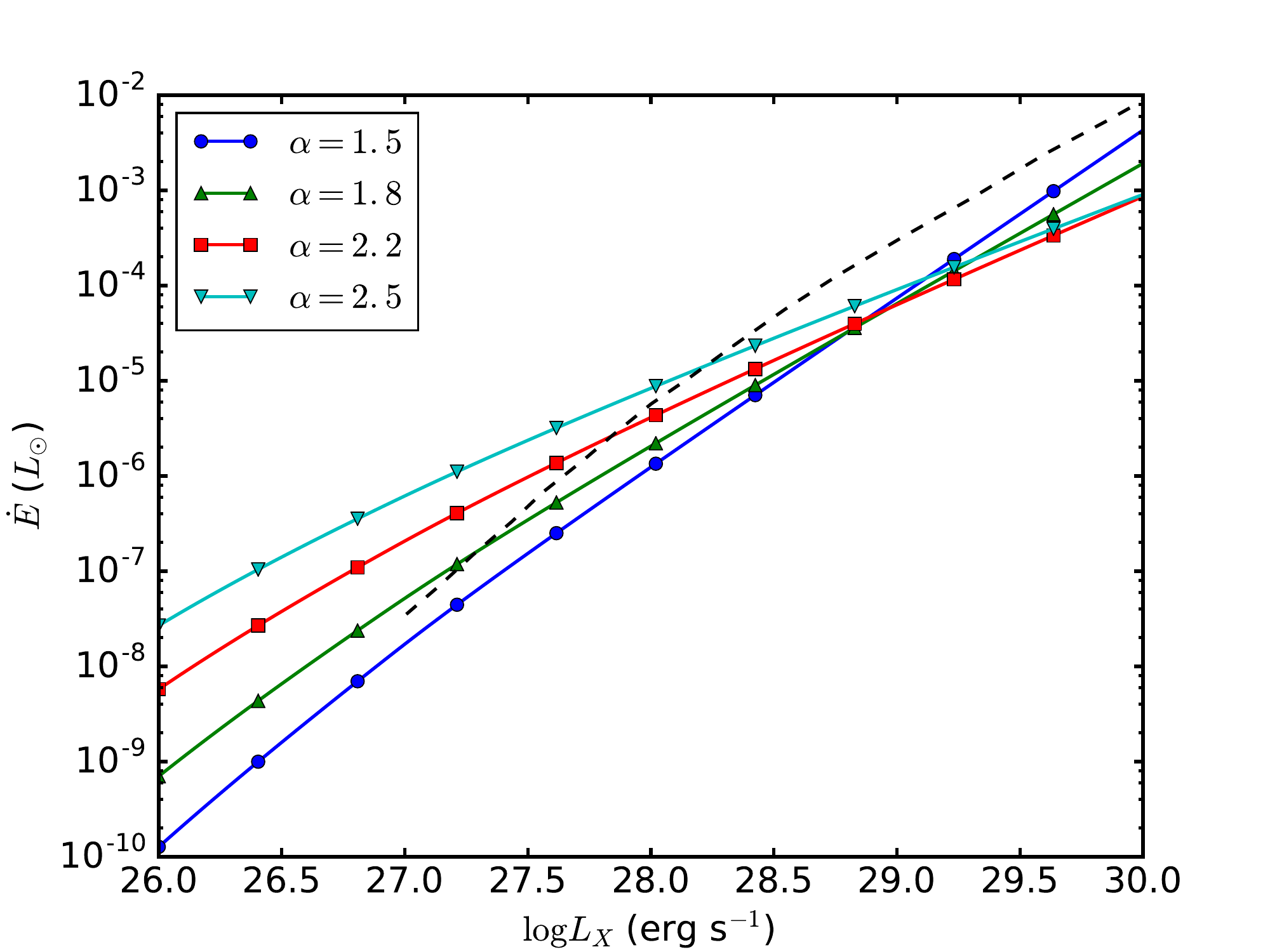}
\caption{Kinetic energy loss rates as a function of $L_X$ and $\alpha$. The relation found by \citet{Drake13} is shown as a dashed line.}
\label{fig:edot}
\end{figure}

We can also estimate the kinetic energy loss rates $\dot{E}_\mathrm{kin}$ due to CMEs. The kinetic energies of solar CMEs also scale with the energies of the corresponding flares
\begin{equation}\label{eq:ekin}
E_\mathrm{kin}=\epsilon E_{GOES}^\delta
\end{equation}
\citep{Drake13}, where $\epsilon=10^{0.81\mp0.85}$ and $\delta=1.05\pm0.03$. The kinetic energy loss rate has the same form as Eqs.~\ref{eq:mdot1} and \ref{eq:mdot2}, but $\mu$ and $\beta$ have to be replaced with $\epsilon$ and $\delta$ in Eq.~\ref{eq:km}. Furthermore, we transform the flare--CME association rate (Eq.~\ref{eq:pm}) $P(M)\rightarrow P(E_\mathrm{kin})$ and find the corresponding parameters $c=-10.25$, $d=0.35$, $E_\mathrm{kin,0}=10^{29}$ and $E_\mathrm{kin,1}=6.8\times10^{31}$\,erg. The resulting kinetic energy loss rates are shown in Fig.~\ref{fig:edot}. To estimate $E_\mathrm{kin,max}$, we use Eqs.~\ref{eq:ekin}, \ref{eq:estar}, and the maximum flare energy estimate from section~\ref{sec:mdotcme}. For the most active stars with $L_X=10^{30}\,\mathrm{erg\,s^{-1}}$, $\dot{E}_\mathrm{kin}$ is about 0.1--0.4~per cent of the solar bolometric luminosity, depending on $\alpha$. This is lower than the value of 1~per cent found by \citet{Drake13} for the same $L_X$. Note also that for the highest $L_X$, $\dot{E}_\mathrm{kin}$ is higher for smaller $\alpha$, in contrast to the behavior of $\dot{M}$.

\subsection{Collision rate with planets}
An important application of such empirical CME models is the estimation of impact rates on orbiting planets, which can inform us about expected ``Space Weather'' effects in exoplanet systems. The impact frequency $f_\mathrm{imp}$ can be written as
\begin{equation}
f_\mathrm{imp}=P_1 P_2 f_\mathrm{CME}
\end{equation}
\citep{Khodachenko07a}, where $f_\mathrm{CME}$ is the intrinsic stellar CME rate and $P_{1,2}$ the meridional and azimuthal impact probabilities, respectively,
\begin{equation}
P_1=\frac{\sin\left[(\Delta+\delta_\mathrm{pl})/2\right]}{\sin\Theta}, P_2=\frac{(\Delta+\delta_\mathrm{pl})}{2\pi},
\end{equation}
where $\Delta_\mathrm{CME}$ is the angular width of the CME, $\delta_\mathrm{pl}$ is the angular width of the planet as seen from the star, and $\Theta$ is the maximum stellar latitude of CMEs (i.e., CME source locations are within $\pm\Theta$). The planet is assumed to orbit in the stellar equatorial plane. All angular quantities are given in radians. Note that $P_1=1$ if $\Theta<(\Delta+\delta_\mathrm{pl})/2$. To estimate the impact rate resulting from the distributions calculated before, we need to transform the mass-dependent occurrence rate (Eq.~\ref{eq:dndm2}) into one depending on angular width. On the Sun, there is a correlation between the apparent angular widths (width projected onto the plane-of-sky) and masses of CMEs \citep{Gopalswamy05, Aarnio11}. We use the relation of \citet{Gopalswamy05},
\begin{equation}\label{eq:mw}
\log M = 12.6+1.3\log \Delta_\mathrm{app},
\end{equation}
where $M$ is in grams and the apparent width $\Delta_\mathrm{app}$ in degrees. This relation was calibrated for CMEs with apparent widths ${\le}120\degr$. The total impact rate is thus
\begin{equation}\label{eq:imp}
f_\mathrm{imp} = \int_{\Delta_0}^{\Delta_\mathrm{max}} P_1 P_2 \frac{dN}{dM} \frac{dM}{d\Delta_\mathrm{app}}\frac{d\Delta_\mathrm{app}}{d\Delta_\mathrm{true}}\ d\Delta_\mathrm{true},
\end{equation}
where $dN/dM$ is taken from Eq.~\ref{eq:dndm2} and $dM/d\Delta_\mathrm{app}$ can be derived from Eq.~\ref{eq:mw}. We introduce a correction factor $d\Delta_\mathrm{app}/d\Delta_\mathrm{true}$, which accounts for the fact that the impact rate depends on the true angular widths of the CMEs. Equation~\ref{eq:mw} relates the masses only with the apparent widths from observations, which are affected by projection effects. Solar CMEs appear systematically larger if they are not propagating in the plane-of-sky and their true widths depend on their source locations and their direction of propagation, so there is no unique transformation between these quantities. Overestimated angular widths would lead to overestimated impact rates. We apply a correction of this effect using the observed and true distributions of solar CME widths \citep{Wu11}. By forward modeling of a large sample of artificial CMEs and applying projection effects, these authors obtained distributions of the apparent CME parameters. Those were then compared with observations to infer the underlying true distribution. We calculate $|d\Delta_\mathrm{app}/d\Delta_\mathrm{true}| = f(\Delta_\mathrm{true})/f(\Delta_\mathrm{app})$ using Eqs.~2 and 9 from \citet{Wu11}. We then integrate Eq.~\ref{eq:imp} between $\Delta_0\approx20\degr$ corresponding to $M_0$ and $\Delta_\mathrm{max}$, which we choose to be $180\degr$, but the inferred impact rates are not very sensitive to the exact upper limit.

\begin{figure}
\centering
\includegraphics[width=\columnwidth]{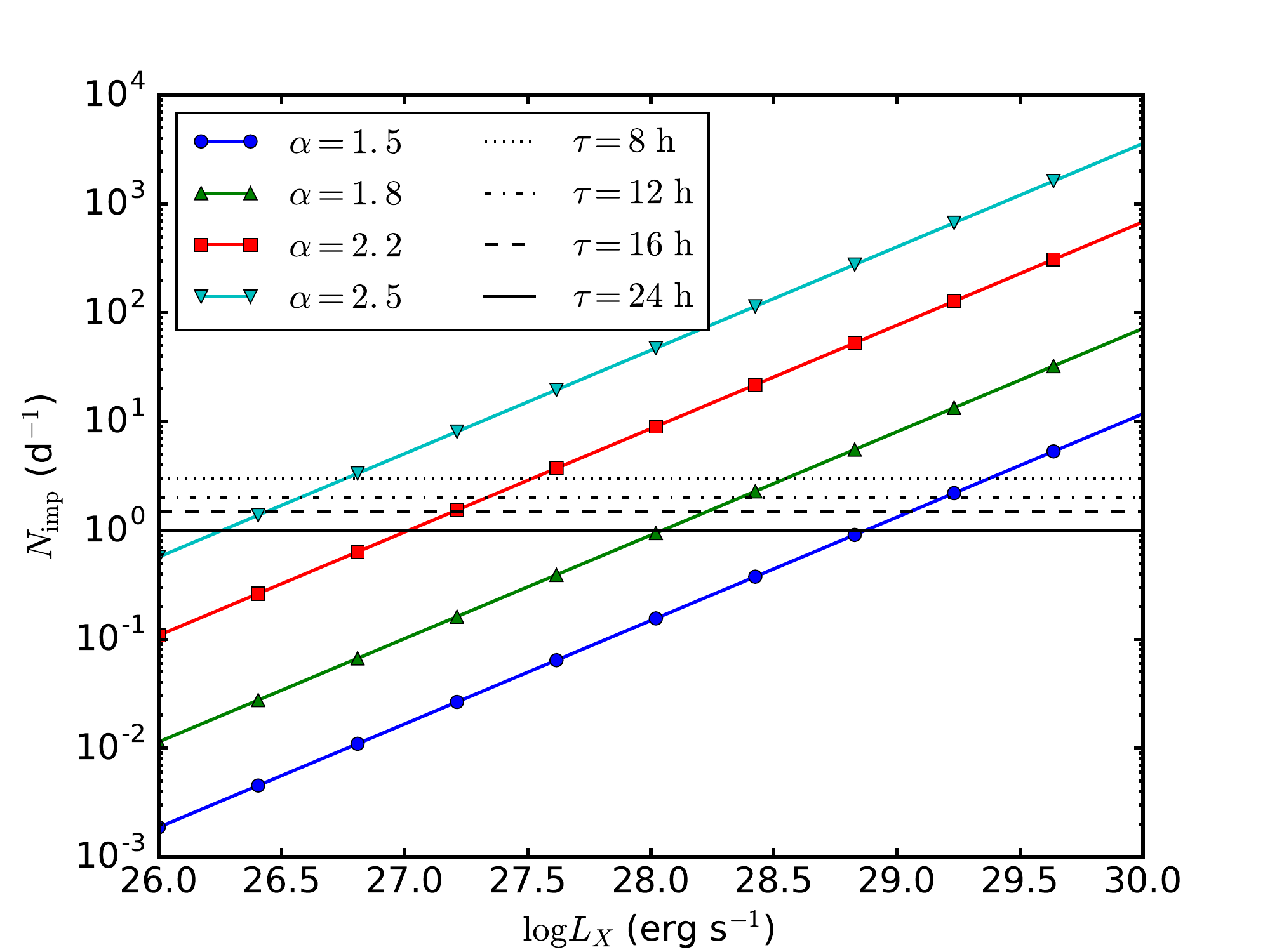}
\caption{CME impact rates on a planet as a function of $L_X$ and $\alpha$. For different CME durations $\tau$, planets orbiting stars with $N_\mathrm{imp}$ above the plotted lines would be permanently exposed to CME plasma.}
\label{fig:imp}
\end{figure}

The resulting CME impact rates on a planet are shown in Fig.~\ref{fig:imp}. As the angular size of a planet is typically very small compared to that of a CME for a wide range of orbital distances, including the location of the HZs of F--M stars, we simply set $\delta_\mathrm{pl} \approx 0$. The impact rates are shown as a function of $L_X$ and $\alpha$. We adopted $\Theta=90\degr$, which may underestimate the impact rate for stars with CME locations preferentially at low latitudes, but overestimates the rates for stars where CMEs occur mainly near the polar regions. However, this choice varies the impact rates by less than a factor of two, which is small compared to the spread of almost three orders of magnitude for different choices of $\alpha$. Note that ignoring the correction factor $d\Delta_\mathrm{app}/d\Delta_\mathrm{true}$ would increase the resulting impact rates by factors of 2--4 for $\alpha$ of 2.5--1.5, independent of $L_X$. For the Sun (adopting $\alpha=1.8$), we find $N_\mathrm{imp}$ of about 10--100 per year for the range of $L_X=2.7\times10^{26}{-}4.7\times10^{27}\mathrm{erg\,s^{-1}}$ from \citet{Peres00}, which is roughly in the order of frontsided Halo CMEs and geoeffective CMEs during the solar cycle \citep{Wang02}. Note that in the solar system about 40--70~per cent of Earth-directed Halo CMEs are geoeffective \citep{Wang02, Gopalswamy07}. Thus, we conclude that the results are reasonable, despite ignoring many details of CME propagation like non-radial motion, deflection \citep[thoroughly discussed in][]{Kay16}, shape, and magnetic field orientation \citep[e.g.][]{Kim08}. Note that the estimated ratio of CME impacts to intrinsic ejections is about 2--4~per cent, very similar to the ratio of Halo CMEs to all CMEs of 3.6~per cent on the Sun \citep{Gopalswamy07}.

Depending on the typical duration of CMEs, young planets orbiting active stars may be under constant CME exposure, thus mimicking a faster, denser stellar wind that may lead to increased ion pick-up loss rates \citep{Khodachenko07, Khodachenko07a, Kislyakova13a, Kislyakova14}. This occurs if $N_\mathrm{imp}\tau{\ge}1$, where $\tau$ is the typical duration of a CME \citep{Khodachenko07a}. For the Sun, $\tau{\approx}8$\,h at 6--10\,R$_{\sun}$, but this number can either increase or decrease with distance from the Sun \citep{Lara04}. It is a lower limit, since it is only measured for the leading edge. In comparison, the durations of magnetic clouds at 1\,au are on average 19--22\,h, with maximum values up to 48\,h \citep{Gopalswamy15d}. However, at 1\,au they already have dimensions of about 0.2\,au, which explains their longer duration. To fully recover from the passage of a CME, the interplanetary medium needs about 2--5\,d \citep{Temmer17}. We plot limits for different $\tau$ from 8 to 24\,h in Fig.~\ref{fig:imp}, because cooler stars have closer HZs and possibly shorter $\tau$ than magnetic clouds (MCs) at 1\,au. For $N_\mathrm{imp}{>}1/\tau$, a planet in the HZ would be exposed to a permanent CME wind. \citet{Khodachenko07a} estimated ion pick-up loss rates for a ``Hot Jupiter'' ranging from $1.5\times10^{11}$ to $9.5\times10^{16}\,\mathrm{g\,s^{-1}}$, depending on the CME density and velocity at the planet's orbit, as well as the planetary magnetic moment, assuming such permanent CME exposure. In the HZs of M dwarfs ($\sim$0.2\,au), oxygen ion loss rates from CO$_2$-rich planets of up to several tens of bar per Gyr for non-magnetized, and a few bar per Gyr for magnetized planets were estimated \citep{Lammer07a}. For a closer distance of 0.05\,au, the loss rates increase up to $10^4$~bar per Gyr for the non-magnetized, and a few 100~bar per Gyr for the magnetized case. This demonstrates the importance of better understanding stellar CME activity, because of the huge potential impact on planetary atmosphere evolution.

\section{Comparison with observations}\label{sec:comp}
For applying the empirical model to younger and more active stars, the question remains if there is a limit where the predicted results may not be realistic anymore. Since application to active stars requires extrapolation of the solar relations to higher flare energies/CME masses and to stars becoming increasingly non-Sun-like (the same holds for extrapolation to other spectral types, e.g. M dwarfs), it is possible that at some point the adopted scalings may not be valid anymore. Here we attempt to find some observational constraints which could be used to test the predictions of our model.

\subsection{Testing the model with solar observations}
First, we compare the model results with solar observations. Since the model also includes relations derived from active stars \citep{Audard00}, it could be possible that our model does not properly predict CME rates of older, rather inactive stars, like the Sun itself. However, as shown before (Fig.~\ref{fig:dndm}), the observed solar CME distribution \citep{Aarnio12} can be well reproduced with our model if choosing $\alpha=1.8$ and $\log L_X=26.5\,\mathrm{erg\,s^{-1}}$. Using the minimum-to-maximum range of the solar X-ray luminosity of $2.7\times10^{26}$ to $4.7\times10^{27}\mathrm{erg\,s^{-1}}$ \citep{Peres00} together with Eq.~\ref{eq:ncme} to estimate the number of CMEs per day yields 0.7 and 13 for solar minimum and maximum, respectively (for $\alpha=1.8$). Our estimate for the maximum is about a factor of 2--3 higher than the average daily rates from observations in the course of the solar cycle \citep[depending on either manual or automated detection techniques;][]{Yashiro08a}, but on individual days such high numbers have been observed \citep{Gopalswamy09, Bilenko14}. Since the adopted solar values of $L_X$ also correspond to short snapshots of the solar emission rather than long-term averages, we find that the agreement is reasonable. Using Eq.~\ref{eq:mdot1}, we find a CME mass-loss rate of $1.3\times10^{-16}$ to $5.5\times10^{-15}\,\mathrm{M_{\sun}\,yr^{-1}}$, i.e. about 0.7 to 27~per cent of the total solar mass-loss rate \citep[$\dot{M}_{\sun}{\sim}2\times10^{-14}\,\mathrm{M_{\sun}\,yr^{-1}}$;][]{Wang98}. Our results are in agreement with observations, indicating a contribution of a few per cent \citep{Vourlidas10}, although our maximum value is again higher. For the solar kinetic energy loss rates, we find $3.6\times10^{-9}$ to $6.3\times10^{-7}\,\mathrm{L_{\sun}}$. This corresponds to about 5 to 50~per cent of the coronal radiative energy loss rate $L_X$. The average kinetic energy loss rate from observations is in the order of $10^{-7}\,L{\sun}$ \citep{Vourlidas10}, in agreement with our results.

\subsection{Comparison with observed stellar mass-loss rates}\label{sec:mdot}
Measurements of stellar mass-loss rates provide an upper limit for the mass-loss rates through CMEs, since the latter cannot be higher than the total mass-loss rates from observations. Therefore, comparison of our results with observations can be used to test the extrapolation to active stars. On the Sun, the contribution of CMEs to the total mass-loss is in the order of a few percent \citep{Vourlidas10}. If CMEs occur only sporadically, their contribution to stellar mass-loss will also be negligible, like on the Sun, and mass-loss will be dominated by the steady wind. For very active stars, CMEs may occur so frequently that, as an ensemble, they mimic an almost steady wind driven by continuous eruptions. In this case, the related mass-loss rates could be so high that CMEs may be the dominant source of mass-loss. Theoretical stellar wind models suggest that a young Sun-like star at an age of about 100\,Myr could have wind-driven mass-loss rates of several $10^{-14}$ to $10^{-13}\,\mathrm{M_{\sun}\,yr^{-1}}$ (depending on its rotation rate), i.e. only up to about 10 times higher than the present Sun \citep{Johnstone15a, Johnstone15b}. In comparison, estimated CME mass-loss rates could be $10^{-12}$ to $10^{-11}\,\mathrm{M_{\sun}\,yr^{-1}}$ for such a star (Fig.~\ref{fig:mdot}).

Several attempts to measure mass-loss rates of cool main-sequence stars have been made. Maximum possible rates were estimated by \citet{Lim96}, who argue that observation of non-thermal coronal radio emission in young stars places an upper limit of about $10^{-12}\,\mathrm{M_{\sun}\,yr^{-1}}$ on any ionized circumstellar material. An additional constraint is the fact that any potential emission from a dense wind cannot exceed observed fluxes in X-rays to UV. Moreover, X-ray and EUV observations towards active stars do not indicate abnormal column densities, which provides an additional upper limit to any neutral circumstellar material \citep{vandenOord97}. Thus, mass-loss from active cool stars should be tenuous and thus difficult to detect. These upper limits are comparable to the maximum CME-related mass-loss rates we obtain here for our most active stars. This indicates that it is unlikely that we underestimate the CME rates with our approach, also justifying to ignore potential CMEs that are not related to flares.

Other methods, including the search for free-free emission from fully ionized winds \citep{Lim96b, Gaidos00, Guedel02, Fichtinger17} or charge exchange X-ray emission at the astropause \citep{Wargelin01, Wargelin02}, provided further upper limits of similar order of magnitude. However, the detection of excess absorption in Ly\,$\alpha$ by hot neutral hydrogen, which is produced at the astropause by charge exchange of the stellar wind and the interstellar medium, provided the first actual observations. Mass-loss rates have been derived for about a dozen of cool stars with this method \citep{Wood01, Wood02, Wood05, Wood14}. The presently available Ly\,$\alpha$ data suggest a maximum mass-loss rate of about $100\,\dot{M}_{\sun}$ for Sun-like and cooler main-sequence stars. This maximum is obtained for moderately active stars, however, even more active stars seem to have much smaller mass-loss rates than suggested by the increase from inactive to moderately active stars.

Recently, yet another method to probe tenuous stellar winds has emerged. Close-in exoplanets can act as a probe of the local stellar wind conditions. By modeling the exoplanet's evaporating atmosphere causing excess Ly\,$\alpha$ absorption during transits, \citet{Vidotto17b} determined the stellar wind parameters of the nearby M dwarf GJ~436b, based on \citet{Bourrier16}. They estimated mass-loss rates of 0.5 to $2.5\times10^{-15}\,\mathrm{M_{\sun}\,yr^{-1}}$. As before, the method is not able to distinguish between CMEs and the steady wind, so the inferred mass-loss rates will include both contributions.

\begin{figure*}
	\centering
	\includegraphics[width=\textwidth]{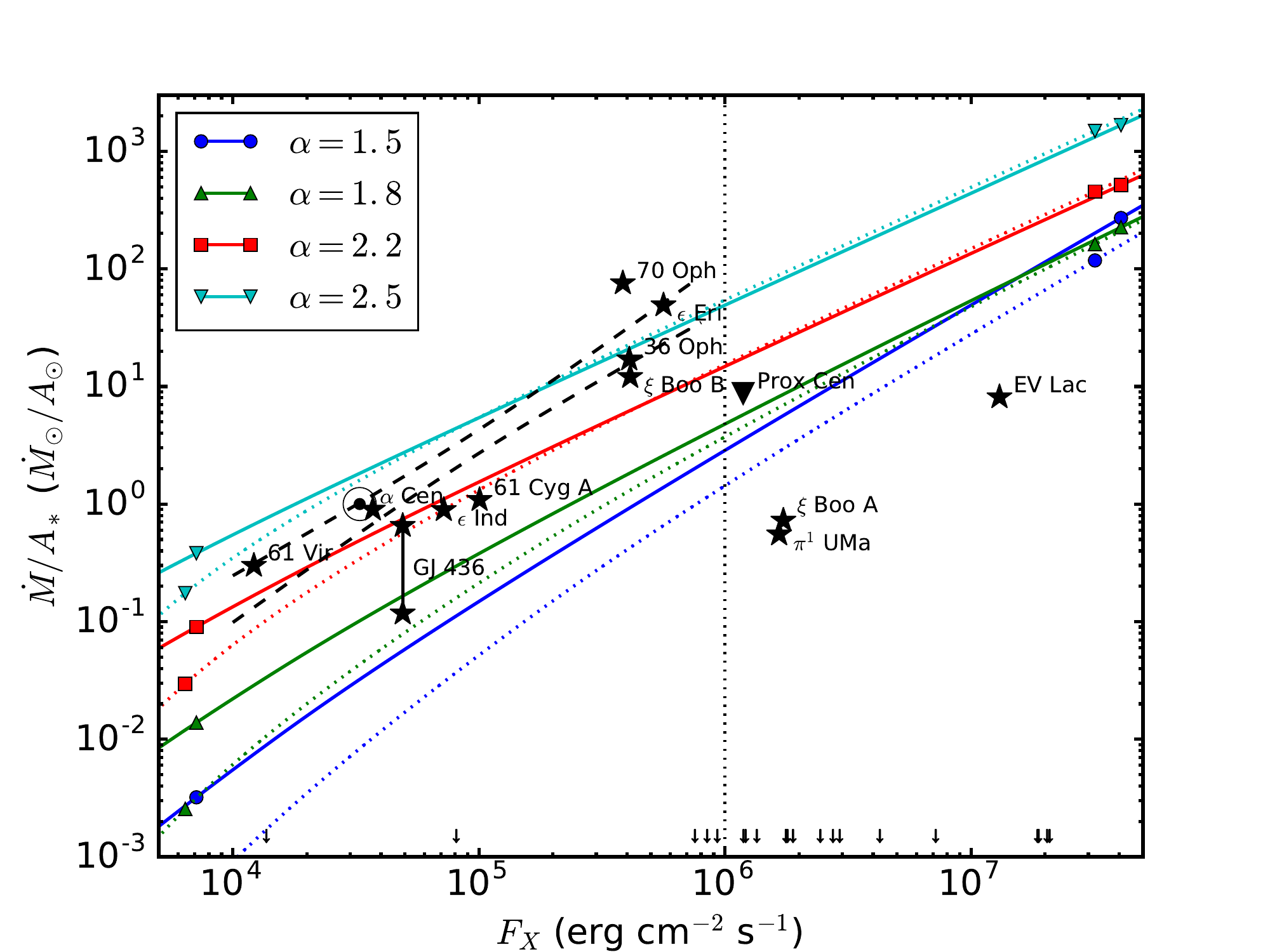}
	\caption{Stellar mass-loss rates normalized to surface area (in solar units) as a function of surface X-ray flux. Stars are data from Ly\,$\alpha$ observations and the down-pointing triangle denotes the upper limit for Proxima Cen \citep{Wood14}. The range of mass-loss rates for GJ~436 is taken from the exoplanet transit modeling of \citet{Vidotto17b}. Arrows along the lower $x$-axis indicate the surface X-ray fluxes of main-sequence F--M stars for which astrospheric Ly\,$\alpha$ absorption could not be detected \citep{Wood05a}. The dashed lines denote the 1$\sigma$ limits to the power-law fit proposed by \citet{Wood05}, the vertical dotted line indicates the approximate cut-off of this scaling. The solid and dotted lines show our modeled CME-related mass-loss rates, where the results are normalized to a Sun-like star (1\,R$_{\sun}$, solid) and to an M dwarf (0.3\,R$_{\sun}$, dotted).}
	\label{fig:wood}
\end{figure*}

In Fig.~\ref{fig:wood}, we reproduce Fig.~4 of \citet{Wood14}, which shows the total mass-loss rate per unit surface area in solar units as a function of surface X-ray flux for the stars with detected Ly\,$\alpha$ absorption. Also shown is the proposed power law fit from \citet{Wood14} of mass-loss rate as a function of $F_X$, valid for inactive to moderately active stars. From their data, there is a clear indication of a cut-off at $F_X{\sim}10^6\,\mathrm{erg\,s^{-1}\,cm^{-2}}$. For more active stars, the observed mass-loss rates are much smaller than extrapolation of the power law would suggest. Furthermore, we include the mass-loss rate of GJ~436 \citep{Vidotto17b}, as derived from modeling the transit observations of its exoplanet. We overplot our results from Eq.~\ref{eq:mdot1} for several $\alpha$ (see also Table~\ref{tab:cmestars}), adopting both a Sun-like star (1\,R$_{\sun}$, solid lines) and an M~dwarf (0.3\,R$_{\sun}$, dotted lines) for normalization. Our estimated CME mass-loss rates (cf. also Table~\ref{tab:cmestars}) are generally comparable or lower than the total mass-loss rates found from Ly\,$\alpha$ absorption, especially for $\alpha{<}2$. However, for $\alpha{>}2$ our predictions are close to or above the observed rates. The agreement with observations of total mass-loss rates breaks down at the power law cut-off, as our model predicts increasing CME mass-loss rates with increasing X-ray flux. Even under the assumption that mass-loss of active stars may be dominated by CMEs, our results are inconsistent with observed mass-loss rates because these are about an order of magnitude lower than our predicted minimum values. In the framework of a steady wind, \citet{Wood14} attempted to explain their results by a change in magnetic field topology and the emergence of polar spots which lead to a more dipole-like field. However, \citet{Vidotto16b} do not find a significant break in the magnetic field parameters at this activity level. Recent model results indicate that high-latitude spots lead to a significant reduction of stellar mass- and angular momentum loss compared with low-latitude spots or even a purely dipolar field \citep{Garraffo15a}. \citet{Linsky14a} also offer a physical interpretation of both the relationship and its break-down. For the most active stars, they suggest that most field lines could be closed and only few open lines would be available from which winds can escape. In models of Alfv\'en wave-driven stellar winds, it is observed that their mass-loss rates saturate for highly active stars due to enhanced radiative losses \citep{Suzuki13a}. However, this does not yet explain why CME-driven mass-loss is also low despite the high flare rates of active stars. We briefly note that the physical basis of a correlation between global X-ray emission and mass-loss, as proposed by \citet{Wood02, Wood05, Wood14}, is debated \citep{Holzwarth07, Cranmer08, Cohen11b}. The latter studies argue that stellar wind is related to open magnetic flux, whereas the global X-ray emission stems predominantly from regions with closed magnetic flux and thus the mass-loss via stellar winds should be independent of activity level. From the similarity of the slopes found from our predicted CME mass-loss rates and the power law from \citet{Wood14}, it is possible that an enhanced contribution of CMEs could be partly responsible for the increasing mass-loss rates with stellar activity from low to moderately active stars. It has been shown that a higher complexity of the magnetic field leads to a reduction of the wind mass-loss rates, as well as the open magnetic flux \citep{Garraffo16}. The comparison with Ly\,$\alpha$ observations suggests that our predictions of CME rates may become unreliable for stars with X-ray luminosities higher than about $6\times10^{28}\,\mathrm{erg\,s^{-1}}$ for G star and $5\times10^{27}\,\mathrm{erg\,s^{-1}}$ for M stars. We caution that this conclusion is based on a very small sample of stars with observed mass-loss rates and the fact that the Ly\,$\alpha$ absorption method relies on detailed knowledge of the ISM parameters towards the stars \citep[e.g., in a fully ionized ISM, no astrospheric Ly\,$\alpha$ absorption can be produced;][]{Guedel14}, which could affect the determined mass-loss rates. However, there is a surprisingly high number of non-detections of astrospheric Ly\,$\alpha$ absorption for many active stars \citep{Wood05a, Wood14}, which we also show in Fig.~\ref{fig:wood} (arrows along the lower $x$-axis). Thus, an exclusively ISM-related effect is unlikely. If the low mass-loss rates of active stars are confirmed by more observations in the future, this would also better constrain predictions of stellar CME rates.

\section{Discussion}\label{sec:disc}
Here we discuss the limitations of our empirical CME model and its application to active stars, including possible modifications of the adopted solar relations at higher stellar activity levels.

\subsection{Flare rates and flare power law index}
The results shown in the previous sections have a rather strong dependence on the choice of the flare power law index $\alpha$. In the most extreme cases, mass-loss rates as well as CME occurrence rates show a spread of about two orders of magnitude for the explored range of $\alpha=1.5$ to 2.5. Thus, constraining $\alpha$ is very important to make reliable predictions of stellar CMEs for particular stars. However, the determination of $\alpha$ can depend on the detailed method and sample sizes \citep{DHuys16, Ryan16}. One thing to consider is that the flare indices found for the largest observed flares, as in the study by \citet{Audard00}, may not be representative for the whole range of flare energies used for the CME rate estimates. As discussed by \citet{Aschwanden07}, a more likely distribution than a power law stems from the avalanche model \citep{Lu93} and includes an exponential factor related to some maximum cut-off energy, which is determined by the maximum amount of free magnetic energy which an active region can generate. The cumulative distribution in such a model reads \mbox{$N(>E) \propto E^{-\alpha+1} \exp(-E/E_\mathrm{cut})$}. Thus, the distribution of the most energetic flares, which are typically those observed on stars due to limited sensitivity, likely falls into the range of exponential decay. This results in measuring steeper slopes which are not representative of the distribution of less energetic flares \citep{Aschwanden07}. Thus, observations covering large flare energy ranges \citep[e.g.][]{Hawley14} would be required to disentangle the power law index and the exponential cutoff. However, in many studies only up to 1--2 orders of magnitude in flare energy were covered \citep{Audard00}.

\subsection{Energy band conversion}
Solar and stellar observations are typically done with very different instruments. All solar scalings used here are based on \textit{GOES} flare energies, whereas the stellar flare study of \citet{Audard00} covered a much broader energy range. Therefore, we use solar flare observations performed both with \textit{GOES} and the \textit{XPS} instrument covering a wider energy band to scale \textit{GOES} flare energies to broad band X-rays \citep{Woods06}. The conversion factor depends on flare energy and is smaller for higher and larger for lower flare energies. It is calibrated in an energy range of $E_{XPS}\approx10^{30}$ to $3\times10^{32}$\,erg or $E_{GOES}\approx10^{28}$ to $7\times10^{30}$\,erg and is higher in this range (45--170) than the constant conversion factor of about 15 used in \citet{Leitzinger14} since it covers 0.1--27\,nm. The constant factor, taken from \citet{Emslie12}, represents the energy radiated by the soft X-ray emitting plasma. It was calculated from plasma parameters obtained from the \textit{GOES} data and covers similar energies ($E_{GOES}\approx4\times10^{28}$ to $5\times10^{30}$\,erg). We can perform a check on the suitability of Eq.~\ref{eq:estar} by applying it to the \textit{GOES} X-ray background luminosity. The solar \textit{GOES} background flux ranges from $10^{-8}\,\mathrm{W\,m^{-2}}$ during minimum to a few $10^{-6}\,\mathrm{W\,m^{-2}}$ in maximum \citep{Veronig03, Winter14}. For instance, a range of $10^{-8}$ to $3\times10^{-6}\,\mathrm{W\,m^{-2}}$ corresponds to luminosities of $2.8\times10^{22}$ to $8.4\times10^{24}\,\mathrm{erg\,s^{-1}}$. Applying Eq.~\ref{eq:estar} yields $6.1\times10^{25}$ to $5.8\times10^{27}\,\mathrm{erg\,s^{-1}}$, in agreement with the range of estimated solar broad band X-ray luminosities in the \textit{ROSAT} band \citep{Peres00, Judge03}. The differences in energy bands of \textit{XPS} and \textit{ROSAT} could yield an overestimate of $E_X$ if assuming $E_X\approx E_{XPS}$, since the \textit{XPS} band is wider. Conversely, this means we would underestimate $E_{GOES}$ for a chosen $E_X$ (cf.~Eq.~\ref{eq:estar}), and thus the corresponding CME mass. This would lead to an underestimate of the resulting mass-loss rate. Therefore, the predicted CME occurrence and mass-loss rates from the present study are also slightly lower than in our previous study \citep{Leitzinger14}. A more sophisticated approach by reconstructing the expected emission in the \textit{ROSAT} band for a large sample of solar flares, as done for instance in \citet{Benz94}, would be desirable for future studies.

Another aspect is the extrapolation of Eq.~\ref{eq:estar} to flare energies larger than in solar observations. A large flare on the active RS~CVn star AR~Lac was observed with Beppo-SAX \citep{Rodono99}. Its energy in the band covered by the Low Energy Concentrator Spectrometers (LECS; 0.1--10\,keV) was $1.6\times10^{35}$\,erg, in that of the Medium Energy Concentrator Spectrometers (MECS; 1.7--10\,keV) it was $8.9\times10^{34}$\,erg. Since MECS covers a similar energy range as \textit{GOES} (1.5--12\,keV) and we roughly assume that the \textit{ROSAT} band is comparable to (LECS\,--\,MECS), this would indicate that $E_{ROSAT}{\sim}E_{GOES}$ at these flare energies. Using Eq.~\ref{eq:estar} we obtain $E_X{\sim}E_{ROSAT}{\sim}7E_{GOES}$. This factor may partly be due to the larger flare energy, but also to the wider band of \textit{XPS} compared to \textit{ROSAT}. For given $E_X$, this would lead to an underestimate of the corresponding CME mass by a factor of three at those energies.

\subsection{CME mass--flare energy relation}
There are several possibilities why the CME scalings derived here may become invalid if extrapolated to more active stars. One point is that the solar relation between CME masses and flare energies may break down for events stronger than seen on the Sun.

\citet{Leitzinger14} briefly compared the mass estimates of stellar mass ejection events and the estimated X-ray energies of their associated flares with the solar scaling. We extend their comparison here. Specifically, we consider the events observed as blue-shifted extra-emissions in Balmer lines on the active M dwarfs AD~Leo \citep{Houdebine90}, AT~Mic \citep{Gunn94a}, V374~Peg \citep{Vida16} and the pre-main sequence star DZ~Cha \citep{Guenther97}. Published estimates of their minimum masses are $7.7\times10^{17}$, ${\sim}10^{15}$, ${\sim}10^{16}$, and $1.4\times10^{18}\ldots7.8\times10^{19}$\,g, respectively. Since no simultaneous X-ray observations of the associated flares are available, we estimate the corresponding X-ray flare energies based on the assumption that the published $U$-band energies are of comparable magnitude \citep[cf.][]{Hawley91}. Doing so, we find $E_X\approx E_U{\sim}2\times10^{32}$\,erg for AD~Leo \citep{Rodono85, Hawley91}, $3\times10^{31}$\,erg for AT~Mic \citep{Gunn94a}, and $1.2\ldots2.5\times10^{35}$\,erg for DZ~Cha \citep{Guenther97}, respectively. The recently published event at V374~Peg has an H\,$\gamma$ flare energy of ${\sim}4\times10^{30}$\,erg, which corresponds to ${\sim}10^{32}$\,erg in X-rays \citep{Butler88}. We add two other events, which are, however, more uncertain in their interpretation as real CME events. \citet{Doyle88a} observed a strong increase in neutral hydrogen column density during a flare on the active M dwarf YZ~CMi, which may be interpreted as a rising filament obscuring parts of the flare region. The estimated mass of this event is $3\times10^{17}$\,g and the estimated X-ray flare energy $E_X{\approx}E_U{\sim}8\times10^{30}$\,erg, since the observed X-ray flare emission was likely partly absorbed by the neutral material \citep{Doyle88a}. The second one is a long-decay flare on the young, active M dwarf AU~Mic, which was interpreted as an eruptive event with a CME mass in the order of $10^{20}$\,g by \citet{Cully94}. However, there is a different model of this strong flare (${\sim}3\times10^{35}$\,erg) including posteruptive energy release, which does not involve a CME \citep{Katsova99}. Other studies which observed stellar CME events (cf. section~\ref{sec:intro}) did not provide mass estimates and had to be excluded.

\begin{figure}
	\centering
	\includegraphics[width=\columnwidth]{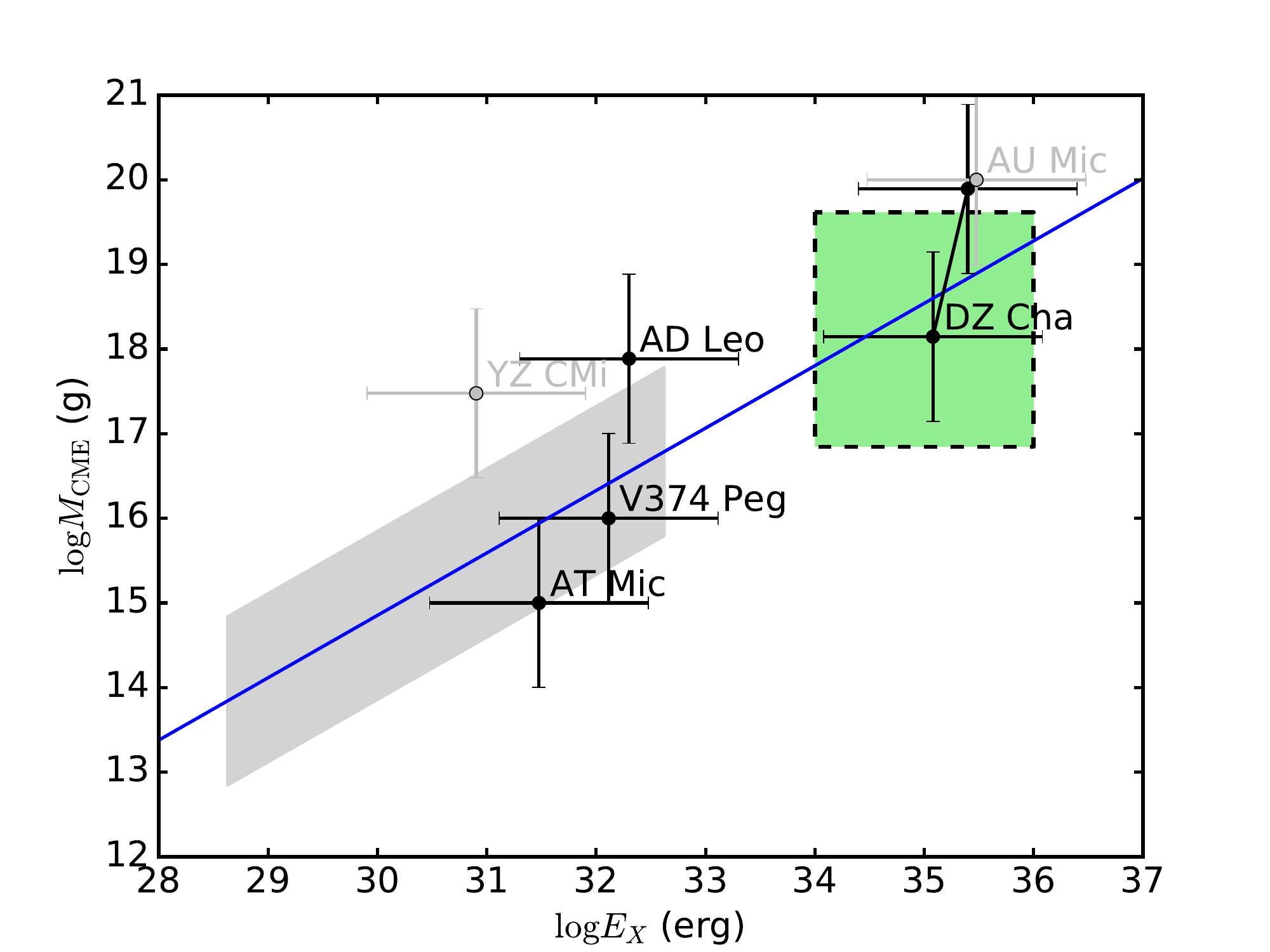}
	\caption{Relation between flare energy and CME masses from Eqs.~\ref{eq:me} and \ref{eq:estar}. The gray shaded area indicates the range of solar data used to calibrate Eq.~\ref{eq:me}, converted to $E_X$ using Eq.~\ref{eq:estar}. Black dots denote observed stellar flare energies and minimum masses of the associated CMEs, gray dots are uncertain events. The square shaded area (dashed outline) shows the range of typical prominence masses of young stars, together with the maximum flare energies observed on them.}
	\label{fig:mex}
\end{figure}

We plot the estimates of stellar CME masses and X-ray flare energies in Fig.~\ref{fig:mex}, in comparison with the empirical CME mass-flare energy scaling from the Sun (Eq.~\ref{eq:me}) converted to broad band X-rays using Eq.~\ref{eq:estar}. For all stellar CME mass and flare energy estimates we plot error bars of an order of magnitude, comparable to quoted uncertainties of the masses. For the X-ray flare energies, this is due to the lack of direct measurements. The extrapolated scaling and the stellar observations agree to within an order of magnitude. Moreover, we show the range of typical masses of stellar prominences of a few ${\sim}10^{16}$ to $10^{19}$\,g \citep{CollierCameron89, Dunstone06a, Leitzinger16}, together with the largest observed X-ray flare energies of these stars. These are in the order of $10^{34}$ to $10^{36}$\,erg for the rapidly rotating, active K dwarfs AB~Dor \citep{Audard00} and Speedy~Mic \citep{Kuerster94}. One can see that this parameter space is also consistent with the extrapolated solar relation. If these large prominences become unstable and erupt, they may lead to CMEs with comparable masses. Moreover, due to energy partition arguments for eruptive events, some proportionality between CME mass and flare emission is very likely \citep{Emslie12, Osten15a}. \citet{Aarnio12} even claim that the solar relation can be extrapolated to T~Tauri stars, which they justified by comparing observed flare energies with estimated post-flare loop masses. The broad agreement, even up to three orders of magnitude above the solar parameter range, suggests that indeed strong flares of active stars may be accompanied by plasma ejections with masses comparable to those predicted by the solar relation. We note again that the observed stellar CME mass estimates are lower limits. Despite being based on a rather small number of events, we conclude that a breakdown or strong deviation of the solar relation at high flare energies is unlikely and not supported by existing observations.

\subsection{Maximum flare energy}
The choice of maximum CME mass corresponding to the maximum flare energy entering Eqs.~\ref{eq:ncme} and \ref{eq:mdot1} has also some impact on the results. It has, however, a negligible impact on the CME rates. If setting $E_\mathrm{max} \approx 10^6 L_X$, i.e. a factor of ten higher, the CME rate for $\alpha=1.5$ is raised by a factor of about 1.4, but for increasing $\alpha$ the difference decreases, reaching zero at $\alpha=2.5$. There is also only a small effect on the CME mass-loss rates. If setting $E_\mathrm{max}$ again to a factor ten higher, the mass-loss rates for $\alpha=1.5$ are raised by a factor of three, but for increasing $\alpha$, the difference decreases down to zero at $\alpha=2.5$. However, the kinetic energy loss rate is affected more strongly. If again setting $E_\mathrm{max} \approx 10^6 L_X$, the kinetic energy loss rates for $\alpha=1.5$ are also raised by about a factor of ten, the increase being slightly higher for low $L_X$. For increasing $\alpha$, the difference in kinetic energy loss rates decreases. For $\alpha=2.5$ and low $L_X$, the difference is about a factor of two, but approaches zero for large $L_X$.

This raises the question of the maximum possible flare energy which a star can produce. This quantity can be estimated as $E_\mathrm{flare}\approx fE_\mathrm{mag} \approx f (B^{2}L^{3})/(8\pi) \approx f (B^{2}A^{3/2})/(8\pi)$ \citep{Aschwanden07, Shibata13}, where $f$ is the fraction of the free magnetic energy $E_\mathrm{mag}$ that can be converted into heating (about 10~per cent), $B$ is the magnetic field strength of the active region producing the flare, and $L$ and $A$ are the active region size and area. The maximum flare energy of the Sun is estimated to be about $6\times10^{33}$\,erg, but it may be in the order of $10^{36}$\,erg for the most active stars \citep{Aulanier13}. Deriving stringent limits for any star is limited by uncertainties in the maximum field strengths of stellar spots and their sizes. Magnetic fields are in the order of a few kG and maximum areas can be up to a few tens of percent of the stellar surface \citep{Berdyugina05}. Determination of stellar spot sizes is difficult because of limited spatial resolution achieved by current Doppler maps and the degeneracy between spot sizes and temperatures. \citet{Jackson13} estimated active region sizes of young M dwarfs by light curve modeling and found length scales between individual regions of about 25\,Mm. Taking this value as a maximum length scale of an individual region and assuming a maximum magnetic field strength of 5\,kG \citep{Berdyugina05}, the maximum flare energy is in the order $10^{33}$\,erg, comparable to broad band optical flare energies \citep{Hawley14}. For stars with transiting exoplanets, spot crossings can be used to infer spot sizes, which yielded spot sizes of 40--150\,Mm for a solar-like G7 star \citep{Silva-Valio10}. This would yield maximum flare energies of $6\times10^{33}$ to $3\times10^{35}$\,erg, the lower value being comparable to that estimated for the Sun.

The maximum attainable flare energy would also be related to a maximum possible CME mass, since also the kinetic energy of CMEs cannot be arbitrarily high. Detailed studies of the energetics of eruptive solar flares found that the total radiated flare energy was about one third of the total energy of the associated CME, which is in turn dominated by its kinetic energy \citep{Emslie04, Emslie05, Emslie12}. A rough estimate of the maximum CME mass can be done by equating the maximum flare energies derived above with the kinetic energy of the CME and assuming a velocity comparable to the stellar escape velocity. This yields a maximum CME mass in the order of $10^{17}$\,g for a flare energy of $10^{33}$\,erg and an escape speed of $600\,\mathrm{km\,s^{-1}}$ for active M dwarfs, comparable to our estimates in section~\ref{sec:mdotcme}. Another important point related to the power law index $\alpha$ is that it also depends on the energy range of the radiation in which the flares are observed \citep{Guedel03}. Our empirical model is related to broad band soft X-ray emission, as frequently used for stellar flare observations, and thus we explore the typical ranges of $\alpha$ in this energy range \citep{Audard00}. Moreover, in the optical, $\alpha$ seems to depend on the effective temperature of the star \citep{Pettersen84}.

\subsection{Association rate}
What could actually reduce the expected high mass- and energy loss rates for the most active stars is a modification of the flare--CME association rate. \citet{Drake13} suggested a possible drop in association rate for the highest flare energies. However, this raises the question on the physical processes generating the most powerful flares on stars. Since on the Sun the probability that a flare is eruptive generally increases with increasing flare energy, the process generating the most powerful flares is likely to be closely connected with eruptive structures. We favor an alternative explanation. It is possible that the flare--CME association rate shifts to higher flare energies, meaning that it could have a similar functional form as on the Sun, but the flare energy where the association becomes about 100~per cent is much higher. In this context, it is necessary to review what is known about the differences of confined and eruptive flares on the Sun \citep[e.g.][]{Janvier15}.

\subsubsection{Solar confined and eruptive flares}
Flares without associated CMEs can be divided into two classes: first, there are truly confined flares which have no large-scale restructuring of magnetic fields and are typically weak, such as simple loop flares or emerging flux flares; second, flares with failed eruptions which are produced like eruptive flares, but the material falls back towards the Sun, or is caught in the overlying magnetic field \citep{Sun15, Mrozek11}. Also partial eruptions have been observed on the Sun, where either only a part of a magnetic structure erupts, or some material of an eruption is partly falling back to the surface \citep{Christian15}. Two magnetic forces are involved in an eruption: the magnetic tension of the overlying coronal field (directed inward) and the magnetic pressure of the flux rope (directed outward). As stated previously, the probability of an association with a CME increases with flare energy \citep{Harrison95, Andrews03, Yashiro05, Yashiro06, Yashiro09}. However, in 2014, one of the largest solar active regions (ARs) since 1990 (AR~12192) did not produce any CME from its core region, despite being an efficient emitter of numerous and strong flares, including many X-class flares \citep{Sun15, Chen15}. Thus, while the increasing association rate for stronger flares may be representative on a statistical basis, some individual active regions seem to deviate from this average picture. Possible reasons for the confined nature of strong flares can be either weak non-potentiality of the active region and/or strength and decay of the background magnetic field \citep{Sun15, Thalmann15}.

Comparative studies of confined and eruptive flares of similar magnitude also identified additional aspects. Confined flares are often located much closer to the active region centers, whereas eruptive flares occur closer to the borders \citep{Wang07, Cheng11}. This may, however, be directly related to conditions in the overlying field, since \citet{Cheng11} also found that the transverse field was stronger over the active region center than at its borders. These findings agree with previous studies of \citet{Wang07}, who noted that the ratio of low to high coronal magnetic flux above an active region is generally larger for eruptive flares. On the other hand, the magnetic flux of the source region itself was similar for eruptive and confined flares in their study, which were all of similar magnitude. Asymmetric overlying fields seem to confine eruptions better than symmetric ones \citep{Liu09a}. Failed eruptions often occur in the largest active regions with stronger magnetic flux \citep{Ji03, Liu08}. \citet{Liu16b} compared the properties of productive ARs (producing flares and/or CMEs) with inert ARs and found that productive ARs are always large and have sufficient free energy, but an eruptive region requires additionally a mature sheared or twisted core field or a weak overlying field.

Also theoretical studies support the picture that a steep decrease of the overlying field is necessary to produce successful eruptions. The onset of torus instability occurs only if the decay index of the overlying field is larger than about 1.5 \citep{Toeroek05, Zuccarello15}. Observations of a filament which underwent five failed eruptions and only (partly) erupted at the sixth attempt showed that for almost similar coronal confinement in all cases, the eruption was associated with the most powerful flare which was able to supply more kinetic energy to the filament as in the failed events \citep{Shen11}. This supports the picture that for almost constant conditions of the overlying field, the eruptiveness mainly depends on flare energy. During the solar cycle, the CME rate is generally correlated with the sunspot number, i.e. the number of active regions, but is also affected by the strength and structure of the global magnetic field \citep{Bilenko14}. A weakened polar field emerging in cycle~24 is the likely cause of an increased CME rate per sunspot number \citep{Petrie13, Petrie15}. The importance of the overlying field related to the solar flare--CME association rate was stressed by \citet{Hassanin16}, who state that the smallest eruptive flares (B) and the strongest confined flares (X) on the Sun have energies which differ by several orders of magnitude. In summary, all these studies indicate that both the strength of the flare and the overlying coronal magnetic field are key properties determining if a flare event is also associated with a CME.

\subsubsection{Implications for active stars}
From the discussions above it becomes clear that flare strength is only one factor which may indicate the occurrence of a CME. However, in our present model, we use the average flare--CME association rate from the Sun, which is only a function of flare energy. For the Sun, this relation is representative, since it is an average over a large number of flares from many different active regions. Other effects like the typical distribution of overlying magnetic field strengths are implicitly included. If, for instance, a flare of a given energy has a 30~per cent probability to be associated with a CME, this would mean that only in 30~per cent of ARs where such flares occur the overlying field is weak enough to allow an eruption. This poses a problem for extrapolation of Eq.~\ref{eq:pm} to other stars, especially to younger, more active ones. In the present model, extrapolation to more active stars implies that the average confinement remains similar as on the Sun. Based on the stronger large-scale magnetic fields of active stars and their different topologies \citep[e.g.][]{Donati09, Reiners12b}, it is likely that the typical confining fields could also be stronger. Young Sun-like stars seem to have much stronger toroidal field components than older stars, which could potentially inhibit CMEs \citep{Vidotto16b}. Moreover, active regions on young stars are likely larger and/or there are several neighboring regions so that the magnetic fields overlying the flare locations may be highly complex. Using Eq.~\ref{eq:pm} could then lead to an overestimate of the CME occurrence rate and the corresponding mass- and energy loss rates, because the confinement is not properly scaled to more active stars.

Progress in understanding the role of the confining field can be made by further solar observations and theoretical modeling. For instance, \citet{Gibb16} found that increased differential rotation leads to higher non-potentiality of the corona; they concluded that such stars may eject more CMEs. Since differential rotation increases with stellar effective temperature \citep[i.e. with decreasing convection zone depth;][]{Barnes05, CollierCameron07}, this would suggest that young G-type stars could have more eruptive events than M dwarfs with similar rotation rates, which spin more rigidly. Moreover, if solar studies may, besides making progress in understanding the physics behind the eruptions, establish some observational property of flares distinguishing confined and eruptive events, such diagnostic may also be applied to stellar flare observations. For instance, \citet{Gopalswamy09b} noted that several strong solar flares without associated CMEs also lacked metric type III bursts, which are indicators of non-thermal electrons propagating along open field lines. They concluded that in these cases the electrons could not access open field lines because of strong confinement. Further investigation of this finding could be potentially useful for stellar studies. \citet{Wang16} found that confined flares have a stronger late-phase EUV emission\footnote{Secondary enhancements in some EUV lines after the main flare, sometimes producing more emission than the main flare itself in these lines \citep{Woods11}.} than eruptive flares of similar magnitude, suggesting that the energy of the flux rope is transformed largely to thermal energy if it cannot escape. However, because of the difficulties to observe stellar EUV emission due to strong absorption by the interstellar medium at these wavelengths, this diagnostic is probably not very useful.

Recently, \citet{Harra16} studied a sample of energetic solar flares, both confined and eruptive, with the aim to find some distinctive properties. The only feature they could identify is that eruptive events show dimmings in coronal lines. On the Sun, depth and slope of these dimming events are closely related with CME mass and speed \citep{Mason16}. Future studies are needed to evaluate the applicability of this method to stellar observations. Of course, direct detection of stellar CMEs and constructing observational flare--CME relationships would be the most straightforward way to infer the intrinsic CME rates, but this requires long-term monitoring to catch sufficient events to reach statistically significant results \citep[][Odert et al., in prep.]{Leitzinger14, Korhonen16}. Besides that, the method is restricted to the most massive events that can still be observed in integrated light, which are likely rarer than events of moderate magnitude. If the intrinsic CME rates of active stars are indeed lower than expected due to strong confinement, the latter approach would become increasingly difficult. However, if flares on active stars are predominantly confined, but in the sense of triggering failed filament eruptions, these failed eruptions may be observable with dedicated monitoring. For instance, \citet{Vida16} found, besides a mass ejection during a complex flare event on the young M dwarf V374~Peg, indications of some material falling back towards the star after the event, as well as some slowly rising material prior to the mass ejection, which they interpreted as failed eruption.

\subsubsection{Estimating the effect of coronal confinement}
Estimating how the changing coronal confinement would impact the flare--CME association rate, and therefore the CME occurrence rate, is difficult. To make some crude estimate, we consider a simplified toy model with an association rate described by a step function. Above a certain flare energy $E_\mathrm{crit}$, all flares should be accompanied by CMEs and below, all flares should be confined. This is equivalent of assuming a single value for the typical overlying field strength, so that the eruptiveness of a given flare is only a function of its energy for a given star. According to \citet{Wang07}, the ratio of low to high coronal magnetic flux $\Phi_\mathrm{low}/\Phi_\mathrm{high}$ above an AR must exceed a critical value for a successful eruption. We assume that this should be similar on other stars, if eruptive flares are produced by the same physical process. Then eruptive events require $\Phi_\mathrm{low,*}/\Phi_\mathrm{high,*} \ge \Phi_\mathrm{low,\sun}/\Phi_\mathrm{high,\sun}$, where the latter ratio denotes the critical value from the Sun. Rewriting this expression yields  $\Phi_\mathrm{low,*}/\Phi_\mathrm{low,\sun} \ge \Phi_\mathrm{high,*}/\Phi_\mathrm{high,\sun}$. One can see that if the confining field in the higher corona is stronger than on the Sun, the magnetic flux at lower heights must increase accordingly.

For simplicity, we assume that the overlying field is well represented by the global coronal magnetic field of the star. This neglects possible local enhancements due to the presence of nearby ARs, as well as the flare-producing AR itself. The magnetic field strength at a distance $r$ from the center of the star is given by
\begin{equation}
	B(r) = B_0\left(\frac{R_*}{r}\right)^n,
\end{equation}
where $B_0$ is the photospheric magnetic field strength and $n=3$ for a dipolar field. The photospheric magnetic flux $\Phi_0=4\pi R_*^2B_0$ is well correlated with X-ray luminosity for both the Sun and other stars \citep{Pevtsov03, Reiners12b}, with a scaling law $L_X \propto \Phi^{1.15}$. This results in
\begin{equation}
\frac{\Phi_\mathrm{high,*}}{\Phi_\mathrm{high,\sun}} = \left(\frac{L_X}{L_{X,\sun}}\right)^{0.87} \left(\frac{R_*}{\mathrm{R}_{\sun}}\right)^{n},
\end{equation}
where we have assumed the same global magnetic field topology ($n$) for the Sun and the star, as well as the same height $r$ where $\Phi_\mathrm{high}$ is evaluated. In case of a Sun-like star ($R_*=\mathrm{R}_{\sun}$), the overlying field would increase as $L_X^{0.87}$ in this simple model. Therefore, it would be about a factor of $10^3$ stronger for a star with $L_X=10^{30}$\,erg\,s$^{-1}$ if adopting $L_{X,\sun}=10^{26.5}$\,erg\,s$^{-1}$.

We further assume that the critical flare energy is related to the lower coronal magnetic flux, which is dominated by the properties of the source region \citep{Wang07}. The radiated energy in a flare is typically a fraction $f{\sim}0.1$ of the free magnetic energy $E_\mathrm{mag}=B_\mathrm{AR}^2L^3/(8\pi)=\Phi_\mathrm{AR}^2/(8\pi L)$ \citep[e.g.][]{Shibata13}, where $B_\mathrm{AR}$ is the magnetic field strength of the AR, $L$ its length scale, and $\Phi_\mathrm{AR}$ the magnetic flux of the AR. If we estimate $E_\mathrm{crit,*}/E_\mathrm{crit,\sun}=E_\mathrm{mag,*}/E_\mathrm{mag,\sun}$ (which assumes same $f$ for the critical energy) we obtain
\begin{equation}
\frac{E_\mathrm{crit,*}}{E_\mathrm{crit,\sun}} = \left(\frac{\Phi_\mathrm{AR,*}}{\Phi_\mathrm{AR,\sun}}\right)^2,
\end{equation}
if assuming similar length scales $L$. Finally, using $\Phi_\mathrm{AR,*}/\Phi_\mathrm{AR,\sun}$ as a proxy for $\Phi_\mathrm{low,*}/\Phi_\mathrm{low,\sun}$ results in
\begin{equation}
\frac{E_\mathrm{crit,*}}{E_\mathrm{crit,\sun}} = \left(\frac{L_X}{L_{X,\sun}}\right)^{1.74} \left(\frac{R_*}{\mathrm{R}_{\sun}}\right)^{2n}.
\end{equation}
Thus, for a Sun-like star ($R_*=\mathrm{R}_{\sun}$), the critical flare energy for an eruptive flare could be a factor $10^6$ higher for an active star with $L_X=10^{30}$\,erg\,s$^{-1}$ compared to the Sun, indicating that only the most energetic superflares may be accompanied by CMEs on young solar-like stars. On the other hand, \citet{Aschwanden14a} found from observations that the dissipated energy in solar flares scales as $E_\mathrm{flare}\propto B_\mathrm{AR}L^{3/2}$. This would result in a shallower increase of the critical flare energy $\propto L_X^{0.87}$, indicating that $E_\mathrm{crit}$ would be a factor of $10^3$ higher for a star with $L_X=10^{30}$\,erg\,s$^{-1}$.

\begin{figure}
	\centering
	\includegraphics[width=\columnwidth]{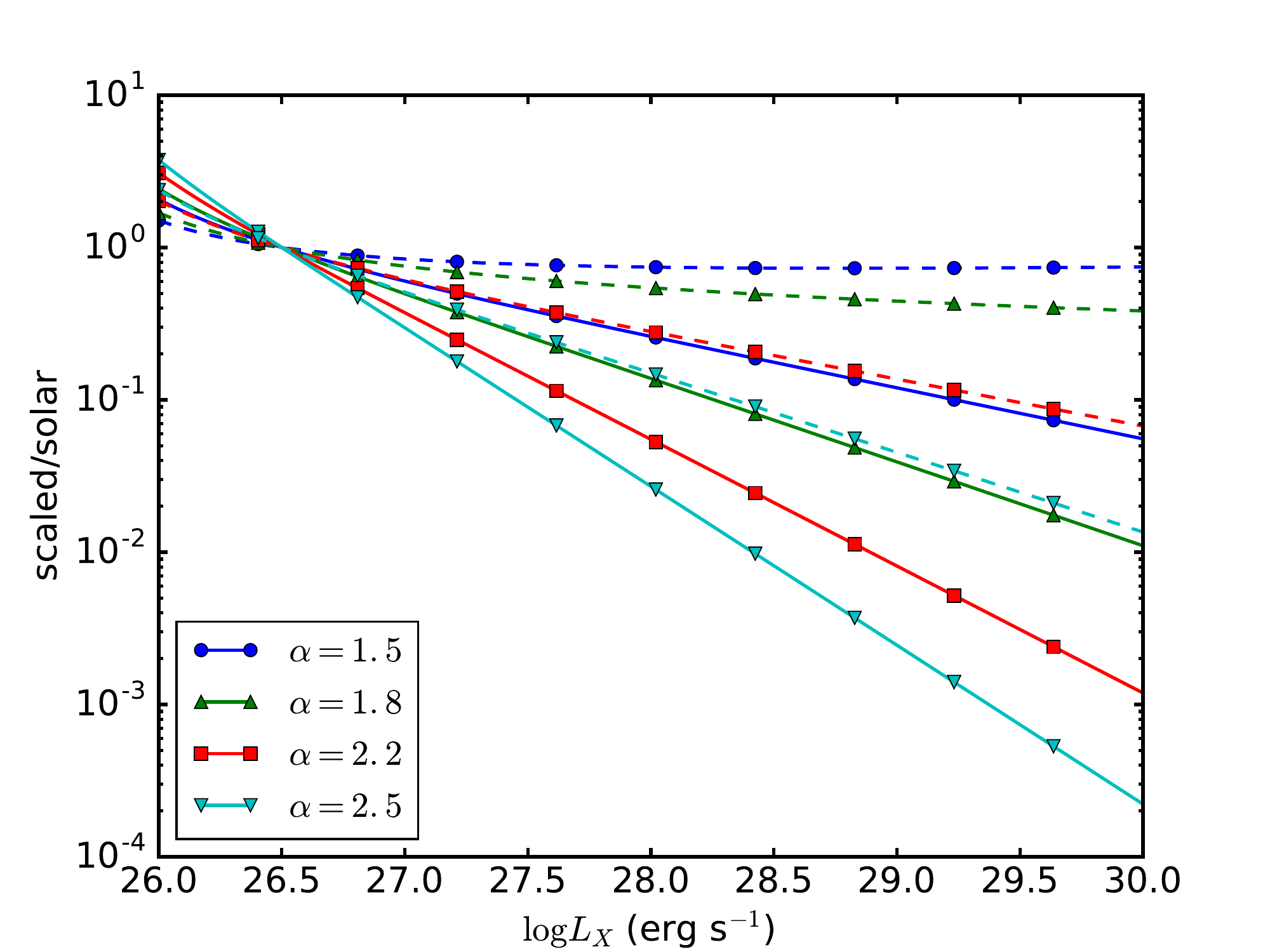}\\
	\caption{Ratio of CME occurrence rates (solid) and mass-loss rates (dashed) calculated with a scaled association rate to ones assuming the solar association rate.}
	\label{fig:assoc}
\end{figure}

We calculate the CME occurrence and mass-loss rates from Eq.~\ref{eq:dndm2} and the simplified step function association rate derived here. In Fig.~\ref{fig:assoc}, we show the ratios of the respective rates using the scaled association to those obtained with assuming the solar value, for a scaling $\propto L_X^{0.87}$. Note that it was necessary to increase $M_\mathrm{max}$, since for the simplified step function association rate, the critical CME mass corresponding to $E_\mathrm{crit}$ exceeded $M_\mathrm{max}$ for stars with low $L_X$. However, if $M_\mathrm{max}$ is chosen sufficiently high to avoid this, the resulting ratios are not very sensitive to the exact choice of $M_\mathrm{max}$. One can see that the occurrence rates are affected more severely by a scaled association rate than the associated mass-loss. For large $\alpha$, the occurrence rates can be suppressed by several orders of magnitude for the most active stars. For small $\alpha$, they can be lower by factors of 10 to 100. This is similar to the suppression of the mass-loss rates for large $\alpha$, whereas the mass-loss rates for small $\alpha$ are barely affected. Note also that for stars less active than the Sun, the occurrence and mass-loss rates could be higher by factors of a few due to weaker confinement. Assuming the steeper scaling $\propto L_X^{1.74}$ suppresses both rates by even larger factors for active stars.

Although the association rate scaling derived in this section is likely too crude to draw firm conclusions on this issue, the results indicate that this topic should be studied in more detail in the future. What is of importance is that the critical flare energy, which is in reality rather a typical flare energy above which mostly eruptive flares occur, could possibly become higher than the maximum flare energy which can be generated by a star. This could indicate that only the largest superflares are typically accompanied by CMEs, or even that CMEs are rare also in these cases. Note that in reality, active regions on young stars can be much larger than solar ones, increasing $\Phi_\mathrm{flare}$, and thus flares may remain eruptive also for increased confinement. Moreover, due to possibly larger heights of stellar flare regions, the overlying field may be weaker than assumed here, but this depends also strongly on the dominating field topology. Additionally, the overlying field could still be dominated by the AR or neighboring ARs, so that the global field underestimates the local field strength. It is obvious that more detailed knowledge on AR parameters on active stars would be needed to address this problem realistically. However, the toy model indicates that ignoring the modified confinement conditions in active stars could lead to a significant overestimate of associated CMEs if simply assuming that all energetic flares on such stars are eruptive events. Further modeling of flux ropes within realistic stellar magnetic fields \citep[e.g.][]{Drake16} could give important insights on this issue.

\section{Summary}\label{sec:sum}
We develop an empirical model relating stellar flare rates with solar flare--CME relationships to estimate the CME occurrence rates and associated mass-loss on active stars. The mass-loss rates found for the most active stars are higher than existing observations of their total mass-loss rates, indicating limits to the extrapolation to the youngest stars. By evaluating all ingredients entering our empirical model, we identify the solar flare--CME association rate as the most uncertain component. Despite their close relation to flares, CMEs may only occur under certain conditions of the overlying magnetic field. Although flare rates and their maximum energies increase with active region sizes and field strengths, as well as the star's overall activity level, the global magnetic field may hinder eruptions. Since the flare--CME association rate is also dependent on coronal confinement, this is likely the main pitfall of models aiming to predict stellar CME rates simply from flare occurrences. More direct detections of CMEs on active stars, as well as better measurements of total stellar mass-loss rates, would be needed to infer better constraints on the CME rates. Moreover, progress in modeling active region magnetic fields on stars with different global magnetic field topologies and strengths could also provide important insights. We conclude that due to this issue the true CME occurrence rates of active stars, despite their strong flaring, still remain elusive.

\section*{Acknowledgements}
This work was supported by the Austrian Science Fund (FWF) project P22950-N16. PO and HL acknowledge FWF project P27256-N27. AH acknowledges FWF project P27765-N2. We thank the anonymous referee for helpful comments.



\bibliographystyle{mnras}
\bibliography{odert-et-al_revised} 




\appendix

\section{Table}
\begin{table*}
	\caption{Stars with mass-loss measurements from Ly\,$\alpha$ absorption. Spectral types, X-ray luminosities and mass-loss rates are taken from \citet{Wood05, Wood14}. Flare rates ($E_X>10^{32}$\,erg) are calculated using the flare power law from \citet{Audard00}. CME mass-loss rates are calculated as described in  section~\ref{lossrates} for $\alpha=1.8$, the values in brackets give the range $\alpha=1.5{-}2.5$. The last column estimates the ratio of mass-loss rates from CMEs to those of winds (only given for $\alpha=1.8$), assuming $\dot{M}_\mathrm{wind}=\dot{M}_\mathrm{obs}-\dot{M}_\mathrm{CME}$, for cases where $\dot{M}_\mathrm{obs}>\dot{M}_\mathrm{CME}$.} 
	\label{tab:cmestars}
	\begin{tabular}{lcccccc}
		\hline
		star           & spectral & $\log L_\mathrm{X}$ &  flare rate  & $\dot{M}_\mathrm{CME}$ & $\dot{M}_\mathrm{obs}$ & $\dot{M}_\mathrm{CME}/\dot{M}_\mathrm{wind}$ \\
		&   type   &   [erg\,s$^{-1}$]    & [day$^{-1}$] &  [$\dot{M}_{\odot}$]   &  [$\dot{M}_{\odot}$] &   \\ \hline
		Proxima Cen    &  M5.5V          &        27.22        &     0.14     &   0.081 (0.025--1.5)   &         $<$0.2   & $>$0.68    \\
		$\alpha$ Cen   & G2V+K0V         &        27.70        &     0.41     &    0.3 (0.11--4.5)     &           2     & 0.18       \\
		$\epsilon$ Eri &   K1V           &        28.32        &     1.6      &     1.5 (0.74--18)     &           30     & 0.05       \\
		61 Cyg A       &   K5V           &        27.45        &     0.24     &   0.15 (0.052--2.6)    &          0.5    & 0.43       \\
		$\epsilon$ Ind &   K5V           &        27.39        &     0.21     &   0.13 (0.043--2.2)    &          0.5    & 0.35       \\
		36 Oph         & K1V+K1V         &        28.34        &     1.67     &     1.6 (0.78--19)     &           15     & 0.12      \\
		EV Lac         &  M3.5V          &        28.99        &     6.93     &     7.9 (5.2--77)      &           1      & --        \\
		70 Oph         & K0V+K5V         &        28.49        &     2.32     &     2.3 (1.2--26)      &          100     & 0.02      \\
		$\xi$ Boo A    &   G8V           &        28.86        &     5.81     &     5.8 (3.6--58)      &        0.5$^a$   & --        \\
		$\xi$ Boo B    &   K4V           &        27.97        &     1.15     &    0.61 (0.26--8.2)    &        4.5$^a$   & 0.16      \\
		61 Vir         &   G5V           &        26.87        &    0.067     &  0.029 (0.0074--0.67)  &          0.3    & 0.11      \\
		$\pi^{1}$ UMa  &  G1.5V     &        28.96        &     6.49     &     7.4 (4.7--72)      &          0.5     & --        \\ \hline
	\end{tabular}
	\\
	$^a$ Total system mass-loss rate is $5\dot{M}_{\odot}$, this splitting is one of the possibilities discussed by \citet{Wood10}.
\end{table*}



\bsp	
\label{lastpage}
\end{document}